\documentclass[11pt,a4paper]{article}
\usepackage{jheppub}
\usepackage[english]{babel}
\usepackage{graphicx}
\usepackage{framed}
\usepackage{amssymb,amsmath,amsbsy}
\usepackage{amsthm}
\usepackage{dsfont}
\usepackage{mathrsfs}

\usepackage{dsfont}
\usepackage{listings}
\usepackage{systeme}
\usepackage{ytableau}
\usepackage{tensor}
\usepackage{tikz}

\usepackage{color}

\makeatletter
\g@addto@macro\bfseries{\boldmath}
\makeatother	

\DeclareMathOperator{\sgn}{sgn}

\newcommand\Tstrut{\rule{0pt}{2.9ex}}         
\newcommand\Bstrut{\rule[-1.2ex]{0pt}{0pt}}   
\newcommand\TBstrut{\Tstrut\Bstrut}
\newenvironment{Eqnarray}%
     {\arraycolsep 0.14em\begin{eqnarray}}{\end{eqnarray}}
\def\beq{\begin{equation}}
\def\eeq{\end{equation}}
\def\beqa{\begin{Eqnarray}}
\def\eeqa{\end{Eqnarray}}
\def\eq#1{eq.~(\ref{#1})}

\def\eqs#1#2{eqs.~(\ref{#1}) and~(\ref{#2})}

\def\eqst#1#2{eqs.~(\ref{#1})--(\ref{#2})}

\def\half{\tfrac{1}{2}}
\def\lsub#1{_{\lower 1.5pt\hbox{$\scriptstyle#1$}}}
\def\vev#1{\left\langle #1\right\rangle}
\def\phm{\phantom{-}}

\title{Tree-level Unitarity in SU(2)$_L\times$U(1)$_Y \times$U(1)$_{Y'}$ Models}

\author[a]{Miguel P.~Bento,}
\author[b]{Howard E.~Haber,}
\author[a]{Jo\~{a}o P.~Silva,}
\affiliation[a]{CFTP, Departamento de F\'{i}sica, Instituto Superior T\'{e}cnico,
Universidade de Lisboa,\\
Avenida Rovisco Pais 1, 1049 Lisboa, Portugal}
\affiliation[b]{Santa Cruz Institute for Particle Physics,
University of California,\\
1156 High Street, Santa Cruz, California 95064, USA}
\emailAdd{miguel.pedra.bento@tecnico.ulisboa.pt}
\emailAdd{haber@scipp.ucsc.edu}
\emailAdd{jpsilva@cftp.ist.utl.pt}

\abstract
{In models with a U(1) gauge extension beyond the Standard Model, one can derive
sum rules for the couplings of the theory that are a consequence of tree-level unitarity. 
In this paper, we provide a comprehensive list of coupling sum rules for
a general SU(2)$_L\times$U(1)$_Y \times$U(1)$_{Y'}$  gauge theory coupled to
an arbitrary set of fermion and scalar multiplets.  These results are of
particular interest for models of dark matter that employ an extended gauge sector
mediated by a new (dark) $Z^\prime$ gauge boson.  For the case of a minimal
extension of the Standard Model with a U(1)$_{Y'}$ gauge boson,
we clarify the definitions of the weak mixing angle and the electroweak $\rho$ parameter.
We demonstrate the utility of a generalized $\rho$ parameter (denoted by $\rho^\prime$)
whose definition naturally follows from the unitarity sum rules developed in this paper.}

\begin{document}
\maketitle

\section{Introduction}
\label{intro}

The requirement that probabilities cannot exceed unity has profound
implications for models of elementary particles and their interactions.  
This constraint is commonly encountered as the requirement that any successful theory must
comply with perturbative unitarity, thus limiting the growth of scattering amplitudes at large energies.
For example, consider a $2 \rightarrow 2$ scattering of fermions, gauge and/or Higgs bosons,
where $s$ is the square of the energy in the center of momentum reference frame.
By imposing ``tree-level unitarity conditions'',
refs.~\cite{LlewellynSmith:1973yud,Cornwall:1973tb,Cornwall:1974km} have shown that
unbroken or spontaneously broken gauge theories are the only theories with vector bosons
that cancel any potential $s^2$ growth of the amplitudes in the large $s$ limit.
Furthermore, demanding the absence of subleading terms that grow like $s$
requires that the tree-level couplings of gauge fields to scalar fields
must arise from gauge-invariant interactions, which in turn imposes various constraints on
such couplings in the form of sum
rules~\cite{Weldon:1984wt,Gunion:1990kf,Brod:2019bro,Bishara:2021buy}.
Similar constraints also arise by considering the allowed tree-level couplings of gauge fields to fermions.
Finally, tree-level amplitudes that behave as $s^0$ at large energies are also constrained, thus imposing
relations among scalar masses and couplings \cite{Weldon:1984wt,Lee:1977yc,Lee:1977eg}.

The consequences of tree-level unitary for models with $N$ Higgs doublets (NHDM) and gauge group
SU(2)$_L \times $U(1)$_Y$ have been studied in detail,
both in the pure scalar sector \cite{Grinstein:2013fia,Bento:2017eti}
and in its couplings to fermions \cite{Bento:2018fmy}. 
Specific applications have appeared for the softly broken, $\mathbb{Z}_2$-symmetric 2HDM
\cite{Ginzburg:2003fe,Grinstein:2015rtl},
for the most general 2HDM \cite{Ginzburg:2005dt,Kanemura:2015ska},
and for all symmetry-constrained versions of the 3HDM \cite{Bento:2022vsb}.

In this paper,
we consider electroweak models with gauge group SU(2)$_L \times $U(1)$_Y \times $U(1)$_{Y'}$
with particular attention given to the most significant sum rules involving
the U(1)$_{Y'}$ gauge boson (denoted by $Z'$).
The idea that the electroweak group could consist of two U(1) factors has a long history.
Moreover, in such models, kinetic mixing of the two U(1) gauge bosons is possible.
Some early references include \cite{Okun:1982xi,Galison:1983pa,Holdom:1985ag,Foot:1991kb},
while a recent phenomenological exploration can be found,
for example, in ref.~\cite{Curtin:2014cca}.
The impact of an extra $Z'$ on the oblique radiative corrections~\cite{Kennedy:1988sn}
has been addressed in ref.~\cite{Holdom:1990xp,Cheng:2022aau},
while general implications of $Z$--$Z'$ mixing are treated, for example, in
\cite{Langacker:1991pg,Babu:1997st,Langacker:2008yv},
under the implicit assumption that $m_{Z'} > m_{Z}$.
In contrast,
a very light gauge boson was considered already in the early 1980s
\cite{Fayet:1980ss,Fayet:1980ad}.
It gained considerable traction as a mediator in a dark sector that includes a candidate for dark matter,
where it is normally known as ``dark photon'';
examples include \cite{An:2014twa,Backovic:2015fnp,Ilten:2018crw}.
Note that the proposed new dark gauge boson has also been called the 
``dark $Z$''  in ref.~\cite{Davoudiasl:2012ag},
or the ``dark $Z^\prime$'' in  \cite{Alves:2013tqa},
etc.
Constraints on such models from neutrino-electron scattering experiments
have been addressed in refs.~\cite{Bilmis:2015lja,Lindner:2018kjo}, and a number of
related studies can be found in refs.~\cite{delAguila:2011yd,Ko:2012hd,Davoudiasl:2013aya,Ko:2013zsa,Arcadi:2018tly}.

In section~\ref{sec:treelevel}, we generalize the results of refs.~\cite{Gunion:1990kf,Bento:2017eti,Bento:2018fmy} to obtain 
sum rule constraints on couplings of a general SU(2)$_L \times $U(1)$_Y \times $U(1)$_{Y'}$ gauge theory of gauge bosons,
fermions and scalars.   We then apply these results to obtain explicit sum rules involving gauge bosons and scalar bosons in
section~\ref{sec:bosons} and additional sum rules that include the couplings of gauge bosons and scalar bosons to fermions in 
section~\ref{sec:fermions}.  We then highlight in section~\ref{sec:generic}
a few of the most useful sum rules in a theory where the scalar sector only
includes scalar eigenstates that are either electrically neutral or singly charged.

In section~\ref{sec:models}, we focus on an  SU(2)$_L \times $U(1)$_Y \times $U(1)$_{Y'}$ model of a dark $Z^\prime$,
under the assumption that the mass of the $Z'$ is less than $m_Z$.
One can provide exact analytical expressions for this model.
Often, the kinetic mixing parameter $\epsilon$ is assumed
to be very small but nonzero,
and an expansion in $\epsilon\ll 1$ is performed.
We demonstrate
that unitarity sum rules applied to this model can serve as important consistency checks on the resulting approximate expressions
obtained for masses and couplings.  In deriving expressions for various observables, we have stressed the importance of the role
of the weak mixing angle and the electroweak $\rho$ parameter, and we advocate definitions that are suitable for the 
model with the extended electroweak gauge group.  Moreover, we show that there exists a new $\rho^\prime$ parameter
(which generalizes the electroweak $\rho$ parameter) that satisfies $\rho^\prime=1$ at tree level in a model that only contains
scalar multiplets with $T=Y=0$ and/or $T=Y=\frac12$ as a consequence of one of the sum rules previously established.\footnote{We denote
the dimension of an SU(2)$_L$ representation by $2T+1$ and the corresponding U(1)$_Y$ hypercharge is normalized such that
the corresponding electric charge is $Q=T_3+Y$.}
Finally, a few brief conclusions are presented in section~\ref{sec:concl}.

\section{\label{sec:treelevel}Tree-level unitarity}

In this section, we consider generic gauge boson, scalar and fermions.
As mentioned in section~\ref{intro},
to preclude $s^2$ amplitude growth,
we assume that the vector bosons arise from some gauge theory \cite{LlewellynSmith:1973yud,Cornwall:1973tb,Cornwall:1974km}.
In this section we do not specify the gauge group. 
Throughout the text,
we use the results of refs.~\cite{Gunion:1990kf,Bento:2017eti,Bento:2018fmy}
and follow their notation,
where the indices $a,b,c,d,e$ refer to vector bosons,
$i,j,k,l$ to scalar bosons,
and $n,m,p$ to fermions.
Summations with the notation $\sum{}'$ are sums over massive states only
(\textit{i.e.}, excluding massless would-be Goldstone modes
and the photon).

Given a gauge theory, one may define the Feynman rules for gauge-gauge and
gauge-scalar vertices as
\begin{itemize}
\item $A_a^\alpha A_b^\beta A_c^\gamma:\, i g_{abc}
\left[ (p_a - p_b)^\gamma g^{\alpha \beta}
+ (p_b-p_c)^\alpha g^{\beta \gamma} + (p_c - p_a)^\beta g^{\gamma \alpha} \right] $\, ,
\item $A_a^\alpha A_b^\beta \phi_i:\, i g_{abi} g^{\alpha \beta}$\, ,
\item $A_a^\alpha \phi_i \phi_j:\, i g_{aij} (p_i-p_j)^\alpha$\, ,
\item $A_a^\alpha A_b^\beta \phi_i \phi_j:\, i g_{abij} g^{\alpha \beta}$\, ,
\end{itemize}
where all the momenta are assumed to be incoming.

We have not provided a Feynman rule for a four-point vector boson vertex.
This is not an issue as unitarity also implies that this rule must be
related to the three-point vertex, and the latter should
satisfy the Jacobi identity. As a further consequence,
the fact that the three-point vertex
has to satisfy the Jacobi identity also entails that
gauge theories are the only consistent theory of vector
bosons,
as shown by the pioneering
work of Llewellyn Smith \cite{LlewellynSmith:1973yud},
Cornwall, Levin and Tiktopoulos \cite{Cornwall:1973tb,Cornwall:1974km},
and later revisited in ref.~\cite{Bento:2017eti}.

Similarly, 
the Feynman rules involving fermions
are
\begin{itemize}
\item $A_a^\alpha \bar{f}_m f_n:\, i \gamma^\alpha \left( g^L_{amn} P_L
+ g^R_{amn} P_R \right)$\, ,
\item $\phi_i \bar{f}_m f_n:\, i \left( g^L_{imn} P_L
+ g^R_{imn} P_R \right)$\, ,
\end{itemize}
where $P_{R,L} = \frac{1}{2}(1\pm \gamma_5)$ are the projectors which
map Dirac fermions into the chiral basis.

As reviewed in appendix E of ref.~\cite{Bento:2017eti}, tree-level unitarity requires that 
the scattering amplitude for any tree-level $2\rightarrow 2$ scattering processes 
cannot grow with the Mandelstam variables $s$ and/or $t$ (after imposing the kinematical
constraint $s+t+u=\sum_i m_i^2$ to eliminate the dependent Mandelstam variable $u$ in favor of
$s$, $t$ and the squared masses of the two incoming and two outgoing particles).  
Consequently, any coefficient of $s$ and/or $t$ raised to a positive power that appears in the scattering
amplitude must vanish.   The conditions obtained by setting these coefficients to zero
yield the coupling constant sum rules given in sections~\ref{sec:treeboson} and
\ref{sec:treefermion}.  The relevant tree-level Feynman diagrams used in obtaining the $2\to 2$ scattering amplitudes
that yield the coupling constant sum rules are explicitly exhibited in appendix E of ref.~\cite{Bento:2017eti} 
and appendix A of ref.~\cite{Bento:2018fmy}.

\subsection{Tree-level unitarity with bosons}
\label{sec:treeboson}

Consider the tree-level Feynman diagrams for the $2 \rightarrow 2$ scattering process
$A_a A_b \rightarrow A_c A_d$ shown in fig.~1 of ref.~\cite{Bento:2017eti}.
Tree-level unitarity yields
\begin{align}
\label{eq:1}
&\sum_e{}' g_{abe}\, g_{cd\bar{e}} \left[ m_e^2 + \frac{(m_a^2 -m_b^2)
	(m_c^2-m_d^2)}{m_e^2}\right]
\nonumber\\[+2mm]
&- \sum_e{}' g_{ade}\, g_{cb\bar{e}}
\left[ m_e^2 + \frac{(m_a^2 -m_d^2) (m_c^2-m_b^2)}{m_e^2}\right]
\nonumber\\[+2mm]
&
-\sum_e g_{ace}\, g_{bd\bar{e}}\left(m_a^2+m_b^2+m_c^2+m_d^2-2 m_e^2\right) =
\sum_k\left(g_{abk}\, g_{cd\bar{k}} - g_{adk}\, g_{bc\bar{k}}\right),
\end{align}
where the prime in $\sum^\prime$ indicates that the sum only runs over massive gauge bosons.
Next, we consider the tree-level Feynman diagrams
for $A_a A_b \rightarrow A_c \phi_i$ shown in fig.~2 of ref.~\cite{Bento:2017eti}. Tree-level unitarity yields
\begin{align}
\label{eq:2}
&\sum_e{}' \left[g_{abe}\, g_{\bar{e}ci} \left[ \frac{m_a^2 -m_b^2+m_e^2}{2
  m_e^2}\right] -
g_{ace}\, g_{\bar{e}bi} \left[ \frac{m_a^2 -m_c^2+m_e^2}{2 m_e^2}\right]
- g_{bce}\, g_{\bar{e}ai}\right]\nonumber\\[+2mm]
&= \sum_k\left(g_{cik}\, g_{ab\bar{k}} - g_{bik}\, g_{ac\bar{k}}\right).
\end{align}
Finally, we consider the tree-level Feynman diagrams for  $A_a A_b \rightarrow \phi_i \phi_j$ 
shown in fig.~3 of ref.~\cite{Bento:2017eti}. Tree-level unitarity yields
\begin{align}
\label{eq:3}
\sum_k g_{aik} g_{b\bar{k}j} - \frac{1}{2} g_{abij}
+ \frac{1}{4} \sum_e{}' \frac{g_{aei}\, g_{\bar{e}bj}}{m_e^2}
- \sum_e \frac{1}{2} g_{abe}\, g_{\bar{e}ij}=0.
\end{align}

\subsection{Tree-level unitarity involving fermions}
\label{sec:treefermion}

Consider the tree-level Feynman diagrams for the $2\to 2$ scattering process $ \bar{f}_m f_n \rightarrow A_a A_b$ shown in fig.~1 of ref.~\cite{Bento:2018fmy}.  Tree-level unitarity yields
%
\begin{align}\label{eq:4}
  &\sum_p \left[m_p\left( g^R_{a\bar{m}p}\, g^L_{b \bar{p} n} + g^R_{b \bar{m} p}\, g^L_{a \bar{p} n} \right)
  - m_m\, g^L_{a \bar{m} p}\, g^L_{b \bar{p} n} - m_n\, g^R_{b \bar{m} p}\, g^R_{a \bar{p} n}\right] \nonumber
  \\[+1mm]
&
+\sum_e{}'\left[ g_{a b e} \left[\frac{m_a^2-m_b^2+m_e^2}{2 m_e^2}\right]
\left( m_n\, g^R_{\bar{e} \bar{m} n}  - m_m\, g^L_{\bar{e} \bar{m} n}\right)\right] =\frac{1}{2}\sum_k
 g_{abk} g^L_{\bar{k} \bar{m} n}\,.
\end{align}
Next, we consider the tree-level Feynman diagrams
for $ \bar{f}_m f_n \rightarrow A_a \phi_i$ shown in fig.~2 of ref.~\cite{Bento:2018fmy}.  Tree-level unitarity yields
\begin{align}\label{eq:5}
&  \sum_e{}' \frac{1}{2 m_e^2} g_{aei} ( m_n\, g^R_{\bar{e} \bar{m} n}  - m_m\, g^L_{\bar{e} \bar{m} n} ) 
- \sum_k  g_{aik}\, g^L_{\bar{k} \bar{m} n} =
\sum_p \left( g^L_{i \bar{m} p}\, g^L_{a \bar{p} n}- g^R_{a \bar{m} p}\, g^L_{i \bar{p} n} \right) \, .
\end{align}
In both eqs.~\eqref{eq:4}--\eqref{eq:5}, a similar rule
can be obtained by exchanging $L \leftrightarrow R$.

\section{\label{sec:bosons}Bosons in SU(2)$_L \times $U(1)$_Y \times $U(1)$_{Y'}$}

In this section we apply the results obtained in section~\ref{sec:treelevel} to an
SU(2)$_L \times $U(1)$_Y \times $U(1)$_{Y'}$ gauge theory,
with a focus on some relations that are most useful.

\subsection{\label{subsec:bosons_rule1}Rule 1}

First, we consider $a=d=W^+$ and $b=c=W^-$ in
eq.~\eqref{eq:1}. Then,
\begin{align}
&\sum_e{} g_{W^+ W^- e}\, g_{W^- W^+ \bar{e}} \, m_e^2 - \sum_e g_{W^+ W^- e}\, g_{W^- W^+ \bar{e}}
\,(4m_W^2 - 2m_e^2) \nonumber \\[+1mm]
&= \sum_k ( g_{W^+ W^- k}\, g_{W^- W^+ \bar{k}} - g_{W^+ W^+ k}\,
g_{W^+ W^+ \bar{k}} ) \, ,
\end{align}
which simplifies to
\begin{align}
& - \sum_e{} g_{W^+ W^- e}\, g_{ W^+ W^- \bar{e}} \, m_e^2 +
\sum_e g_{W^+ W^- e}\, g_{ W^+ W^- \bar{e}}
\,(4m_W^2 - 2m_e^2) \nonumber \\[+1mm]
&= \sum_k ( g_{W^+ W^- k}\, g_{W^- W^+ \bar{k}} - g_{W^+ W^+ k}\,
g_{W^+ W^+ \bar{k}} ) \,.
\end{align}
Since the photon ($\gamma$) is massless, it follows that
\begin{align}
& 4m_W^2 \, g_{W^+ W^- \gamma}^2 + (4m_W^2 - 3m_Z^2) g_{W^+ W^- Z}^2 +
(4m_W^2 -3m_{Z'}^2) g_{W^+ W^- Z'}^2 \nonumber \\[+1mm]
&= \sum_k ( g_{W^+ W^- k}\, g_{W^- W^+ \bar{k}} - g_{W^+ W^+ k}\,
g_{W^+ W^+ \bar{k}} ) \, ,
\end{align}
which yields
\begin{align}\label{eq:sum1}
& 4m_W^2 \, g_{W^+ W^- \gamma}^2 + (4m_W^2 - 3m_Z^2) g_{W^+ W^- Z}^2 +
(4m_W^2 -3m_{Z'}^2) g_{W^+ W^- Z'}^2 \nonumber \\[+1mm]
&= \sum_k g_{W^+ W^- \phi^0_k}^2 - \sum_k g_{W^+ W^+ \phi^{--}_k}\,
g_{W^- W^- \phi^{++}_k} \, .
\end{align}
By analyzing eq.~\eqref{eq:sum1}, one may further specialize this result
by imposing custodial symmetry. Nevertheless, the parameters involved
in the mass diagonalization of the kinetic Lagrangian are more general
than those of the Standard Model (SM).\par
We now examine the case of $a=W^+$, $b=W^-$, $c=d=Z$:
\begin{align}\label{eq:sum2}
& \frac{m_Z^4}{m_W^2} g^2_{W^+ W^- Z} = \sum_k  g_{W^+ W^- \phi^0_k} \, g_{Z Z \phi^0_k}
 -\sum_k g_{W^+ Z \phi^{-}_k}\,
g_{W^- Z \phi^{+}_k} \, ,
\end{align}
which coincides with eq.~(4.2) in ref.~\cite{Gunion:1990kf}.
We note that
the coupling $g_{W^- Z \phi^{+}_k}$ is not found in a multi-Higgs doublet extension of the SM.

We now compute the case of $a=W^+$, $b=W^-$, $c=d=Z'$. Is
is straightforward to see that it is similar to eq.~\eqref{eq:sum2},
\begin{align}\label{eq:sum3}
& \frac{m_{Z'}^4}{m_W^2} g^2_{W^+ W^- Z'} = \sum_k  g_{W^+ W^- \phi^0_k} \, g_{Z' Z' \phi^0_k}
 -\sum_k g_{W^+ Z' \phi^{-}_k}\,
g_{W^- Z' \phi^{+}_k} \, ,
\end{align}
with just an interchange of $Z$ with $Z'$.

Two further relations can be derived.
For $a=W^+$, $b=W^-$, $c=Z$, $d=Z'$:
\begin{align}\label{eq:sum4}
& \frac{m_Z^2 m_{Z'}^2}{m_W^2} g_{W^+ W^- Z'}\, g_{W^+ W^- Z} = \sum_k 
g_{W^+ W^- \phi_k^0}\, g_{Z Z' \phi_k^0} - \sum_k g_{W^+ Z' \phi_k^-}\,
g_{W^- Z \phi_k^+} \, .
\end{align}
For $a=Z$, $b=Z'$, $c=Z'$, $d=Z$:
\begin{align}\label{eq:sum5}
\sum_k g^2_{Z Z' \phi^0_k} = \sum_k g_{Z Z \phi^0_k} g_{Z' Z' \phi^0_k} \, .
\end{align}
Due to the gauge group structure and invariance, couplings of the form $g_{W^+ \gamma \phi_k^-}$
and $g_{\gamma Z Z'}$ are forbidden. Then, the rules eqs.~\eqref{eq:sum1}--\eqref{eq:sum5} 
are the only non-trivial
sum rules for eq.~\eqref{eq:1} with SU(2)$_L \times $U(1)$_Y \times $U(1)$_{Y'}$.

\subsection{\label{subsec:bosons_rule2}Rule 2}

Analogously to what we did for eq.~\eqref{eq:1} in
section~\ref{subsec:bosons_rule1}, we now explore
the sum rules arising from eq.~\eqref{eq:2}. Thus, our first rule
is set with $a=W^-$, $b=W^-$, $c=W^+$, $i=\phi_i^+$:
\begin{align}\label{eq:sum6}
& \frac{3}{2} \left[ g_{W^+ W^- Z}\, g_{Z W^- \phi_i^+} + 
g_{W^+ W^- Z'}\, g_{Z' W^- \phi_i^+} \right] \nonumber \\[+1mm] 
& = \sum_k 
g_{W^+ \phi_i^+ \phi_k^{--}}\, g_{W^- W^- \phi_k^{++}} - 
\sum_k g_{W^- \phi_i^+ \phi_k^0}\,
g_{W^+ W^- \phi_k^0} \, .
\end{align}
For $a=W^+$, $b=W^+$, $c=Z$, $i=\phi_i^{--}$:
\begin{align}\label{eq:sum7}
& g_{W^+ W^- Z}\, g_{W^+ W^+ \phi_i^{--}} 
\left( 2- \frac{m_Z^2}{2m_W^2} \right) \nonumber \\[+1mm] 
& = \sum_k 
g_{Z \phi_i^{--} \phi_k^{++}}\, g_{W^+ W^+ \phi_k^{--}} - 
\sum_k g_{W^+ \phi_i^{--} \phi_k^+}\,
g_{W^+ Z \phi_k^-} \, .
\end{align}
For $a=W^+$, $b=W^+$, $c=Z'$, $i=\phi_i^{--}$:
\begin{align}\label{eq:sum7a}
& g_{W^+ W^- Z'}\, g_{W^+ W^+ \phi_i^{--}} 
\left( 2- \frac{m_{Z'}^2}{2m_W^2} \right) \nonumber \\[+1mm] 
& = \sum_k 
g_{Z' \phi_i^{--} \phi_k^{++}}\, g_{W^+ W^+ \phi_k^{--}} - 
\sum_k g_{W^+ \phi_i^{--} \phi_k^+}\,
g_{W^+ Z' \phi_k^-} \, .
\end{align}
For $a=W^+$, $b=W^-$, $c=Z$, $i=\phi_i^{0}$:
\begin{align}\label{eq:sum8}
& g_{W^+ W^- Z} \left[ \frac{1}{2} g_{Z Z \phi_i^{0}} -
\frac{m_Z^2}{2m_W^2} \, g_{W^+ W^- \phi_i^{0}} \right]
+ \frac{1}{2} g_{W^+ W^- Z'} \, g_{Z Z' \phi_i^{0}} \nonumber \\[+1mm] 
& = \sum_k 
g_{Z \phi_i^{0} \phi_k^{0}}\, g_{W^+ W^- \phi_k^{0}} - 
\sum_k g_{W^- \phi_i^{0} \phi_k^+}\,
g_{W^+ Z \phi_k^-} \, .
\end{align}
For $a=W^+$, $b=W^-$, $c=Z'$, $i=\phi_i^{0}$:
\begin{align}\label{eq:sum9}
& g_{W^+ W^- Z'} \left[ \frac{1}{2} g_{Z' Z' \phi_i^{0}} -
\frac{m_{Z'}^2}{2m_W^2} \, g_{W^+ W^- \phi_i^{0}} \right]
+ \frac{1}{2} g_{W^+ W^- Z} \, g_{Z Z' \phi_i^{0}} \nonumber \\[+1mm] 
& = \sum_k 
g_{Z' \phi_i^{0} \phi_k^{0}}\, g_{W^+ W^- \phi_k^{0}} - 
\sum_k g_{W^- \phi_i^{0} \phi_k^+}\,
g_{W^+ Z' \phi_k^-} \, .
\end{align}
For $a=W^+$, $b=Z$, $c=Z'$, $i=\phi_i^{-}$:
\begin{align}\label{eq:sum10}
& g_{W^+ W^- Z} \, g_{W^+ Z' \phi_i^{-}} \left( 1-\frac{m_Z^2}{2m_W^2} 
\right) - g_{W^+ W^- Z'} \, g_{W^+ Z \phi_i^{-}} 
\left( 1-\frac{m_{Z'}^2}{2m_W^2} \right) \nonumber \\[+1mm] 
& = \sum_k 
g_{Z \phi_i^{-} \phi_k^{+}}\, g_{W^+ Z' \phi_k^{-}} - 
\sum_k g_{Z' \phi_i^{-} \phi_k^+}\,
g_{W^+ Z \phi_k^-} \, .
\end{align}
For $a=Z$, $b=W^-$, $c=Z$, $i=\phi_i^{+}$:
\begin{align}\label{eq:sum11}
& g_{W^+ W^- Z} \, g_{W^- Z \phi_i^{+}} \left( 1+\frac{m_Z^2}{2m_W^2} 
\right)  = \sum_k 
g_{W^- \phi_i^{+} \phi_k^{0}}\, g_{Z Z \phi_k^{0}} - 
\sum_k g_{Z \phi_i^{+} \phi_k^-}\,
g_{Z W^- \phi_k^+} \, .
\end{align}
For $a=Z'$, $b=W^-$, $c=Z'$, $i=\phi_i^{+}$:
\begin{align}\label{eq:sum12}
& g_{W^+ W^- Z'} \, g_{W^- Z' \phi_i^{+}} \left( 1+\frac{m_{Z'}^2}{2m_W^2} 
\right)   = \sum_k 
g_{W^- \phi_i^{+} \phi_k^{0}}\, g_{Z' Z' \phi_k^{0}} - 
\sum_k g_{Z' \phi_i^{+} \phi_k^-}\,
g_{Z' W^- \phi_k^+} \, .
\end{align}
For $a=Z$, $b=W^-$, $c=Z'$, $i=\phi_i^{+}$:
\begin{align}\label{eq:sum13}
& g_{W^+ W^- Z} \, g_{W^- Z' \phi_i^{+}} \left(\frac{m_{Z}^2}{2m_W^2} 
\right) + g_{W^+ W^- Z'} \, g_{W^- Z \phi_i^{+}} \nonumber \\[+1mm] 
& = \sum_k 
g_{W^- \phi_i^{+} \phi_k^{0}}\, g_{Z Z' \phi_k^{0}} - 
\sum_k g_{Z' \phi_i^{+} \phi_k^-}\,
g_{Z W^- \phi_k^+} \, .
\end{align}
For $a=Z'$, $b=W^-$, $c=Z$, $i=\phi_i^{+}$:
\begin{align}\label{eq:sum14}
& g_{W^+ W^- Z'} \, g_{W^- Z \phi_i^{+}} \left(\frac{m_{Z'}^2}{2m_W^2} 
\right) + g_{W^+ W^- Z} \, g_{W^- Z' \phi_i^{+}} \nonumber \\[+1mm] 
& = \sum_k 
g_{W^- \phi_i^{+} \phi_k^{0}}\, g_{Z Z' \phi_k^{0}} - 
\sum_k g_{Z \phi_i^{+} \phi_k^-}\,
g_{Z' W^- \phi_k^+} \, .
\end{align}
For $a=Z$, $b=W^-$, $c=W^+$, $i=\phi_i^{0}$:
\begin{align}\label{eq:sum15}
& - g_{W^+ W^- Z} \, g_{W^+ W^- \phi_i^{0}} \frac{m_Z^2}{m_W^2} + g_{W^+ W^- Z} \, g_{Z Z \phi_i^{0}}
 + g_{W^+ W^- Z'} \, g_{Z Z' \phi_i^{0}} \nonumber \\[+1mm] 
& = \sum_k 
g_{W^+ \phi_i^{0} \phi_k^{-}}\, g_{Z W^- \phi_k^{+}} - 
\sum_k g_{W^- \phi_i^{0} \phi_k^+}\,
g_{Z W^+ \phi_k^-} \, .
\end{align}
For $a=Z'$, $b=W^-$, $c=W^+$, $i=\phi_i^{0}$:
\begin{align}\label{eq:sum16}
& - g_{W^+ W^- Z'} \, g_{W^+ W^- \phi_i^{0}} \frac{m_{Z'}^2}{m_W^2} + g_{W^+ W^- Z} \, g_{Z Z' \phi_i^{0}}
 + g_{W^+ W^- Z'} \, g_{Z' Z' \phi_i^{0}} \nonumber \\[+1mm] 
& = \sum_k 
g_{W^+ \phi_i^{0} \phi_k^{-}}\, g_{Z' W^- \phi_k^{+}} - 
\sum_k g_{W^- \phi_i^{0} \phi_k^+}\,
g_{Z' W^+ \phi_k^-} \, .
\end{align}

\subsection{\label{subsec:bosons_rule3}Rule 3}
The case of the third relation,
arising from the scattering process $A_a A_b \rightarrow \phi_i \phi_j$,
is more intricate than the previous scattering processes. 
There are many possibilities for a given arbitrary model,
and many of them are not very useful for realistic models.
Here, we will argue that, given the lack of experimental
evidence of charged scalars thus far,
we will be more interested in neutral and single charged scalars,
but not fields with electric charge $Q>1$ such as
$\phi^{++}$. Of course, the generalization to such models is straightforward,
albeit with tedious calculations.
Thus, we study the possibility of initial states with total charge $Q=0$.
Then, for $a=W^+$, $b=W^-$, $i=\phi^{-Q}_i$,
$j=\phi^{Q}_j$:
\begin{align}
& \sum_k g_{W^+ \phi^{-Q}_i \phi^{Q-1}_k} \, g_{W^- \phi^{1-Q}_k \phi^{Q}_j} - \frac{1}{2}
g_{W^+ W^- \phi^{-Q}_i \phi^{Q}_j} + \frac{1}{4} \sum_e{}' \frac{g_{W^+ e \phi_i^{-Q}} \, 
g_{\bar{e} W^- \phi^{Q}_j}}{m_e^2} \nonumber \\[+1mm] 
& - \frac{1}{2} \left( g_{W^+ W^- Z} \, g_{Z \phi_i^{-Q} \phi^{Q}_j} + 
g_{W^+ W^- Z'} \, g_{Z' \phi_i^{-Q} \phi^{Q}_j} + g_{W^+ W^- A} \, g_{A \phi_i^{-Q} \phi^{Q}_j} \right) = 0 \,.
\end{align}
For example, if we choose
$Q=0$, then $g_{A \phi_i^0 \phi^{0}_j}=0$.  Likewise, by choosing $Q>2$ it follows that
$g_{W^+ e \phi_i^{-Q}} \, g_{\bar{e} W^- \phi^{Q}_j} = 0$ in light of electric charge conservation.

If we choose $a=Z$, $b=Z$, $i=\phi^{-Q}_i$, $j=\phi^{Q}_j$:
\begin{align}
& \sum_k g_{Z \phi^{-Q}_i \phi^{Q}_k} \, g_{Z \phi^{-Q}_k \phi^{Q}_j} - \frac{1}{2}
g_{Z Z \phi^{-Q}_i \phi^{Q}_j} + \frac{1}{4} \sum_e{}' \frac{g_{Z e \phi_i^{-Q}} \, 
g_{\bar{e} Z \phi^{Q}_j}}{m_e^2} = 0 \, .
\end{align}
We may further specialize this result into $Q=0$ and $i=j$, which yields
\begin{align}
& \sum_k g_{Z \phi^{0}_i \phi^{0}_k} \, g_{Z \phi^{0}_k \phi^{0}_i} - \frac{1}{2}
g_{Z Z \phi^{0}_i \phi^{0}_i} +  \frac{g_{Z Z' \phi_i^{0}} \, 
g_{Z' Z \phi^{0}_i}}{4m_{Z'}^2} = 0 \, .
\end{align}
For $a=Z'$, $b=Z'$, $i=\phi^{-Q}_i$, $j=\phi^{Q}_j$:
\begin{align}
& \sum_k g_{Z' \phi^{-Q}_i \phi^{Q}_k} \, g_{Z' \phi^{-Q}_k \phi^{Q}_j} - \frac{1}{2}
g_{Z' Z' \phi^{-Q}_i \phi^{Q}_j} + \frac{1}{4} \sum_e{}' \frac{g_{Z' e \phi_i^{-Q}} \, 
g_{\bar{e} Z' \phi^{Q}_j}}{m_e^2} = 0 \, .
\end{align}
For $a=Z$, $b=Z'$, $i=\phi^{-Q}_i$, $j=\phi^{Q}_j$:
\begin{align}
& \sum_k g_{Z \phi^{-Q}_i \phi^{Q}_k} \, g_{Z' \phi^{-Q}_k \phi^{Q}_j} - \frac{1}{2}
g_{Z Z' \phi^{-Q}_i \phi^{Q}_j} + \frac{1}{4} \sum_e{}' \frac{g_{Z e \phi_i^{-Q}} \, 
g_{\bar{e} Z' \phi^{Q}_j}}{m_e^2} = 0 \, .
\end{align}
For $a=W^+$, $b=W^+$, $i=\phi^{-}_i$, $j=\phi^{-}_j$:
\begin{align}
& \sum_k g_{W^+ \phi^{-}_i \phi^{0}_k} \, g_{W^+ \phi^{0}_k \phi^{-}_j} - \frac{1}{2}
g_{W^+ W^+ \phi^{-}_i \phi^{-}_j} + \frac{1}{4} \left( \frac{g_{W^+ Z \phi_i^{-}} \, 
g_{Z W^+ \phi^{-}_j}}{m_Z^2} + \frac{g_{W^+ Z' \phi_i^{-}} \, 
g_{Z' W^+ \phi^{-}_j}}{m_{Z'}^2} \right) = 0 \, .
\end{align}

Finally, we have for $a=W^+$, $b=Z$, $i=\phi^{-}_i$, $j=\phi^{0}_j$:
\begin{align}
& \sum_k g_{W^+ \phi^{-}_i \phi^{0}_k} \, g_{Z \phi^{0}_k \phi^{0}_j} - \frac{1}{2}
g_{W^+ Z \phi^{-}_i \phi^{0}_j} + \frac{1}{4} \left( \frac{g_{W^+ Z \phi_i^{-}} \, 
g_{Z Z \phi^{0}_j}}{m_Z^2} + \frac{g_{W^+ Z' \phi_i^{-}} \, 
g_{Z' Z \phi^{0}_j}}{m_{Z'}^2} \right) \nonumber \\[1mm]
& + \frac{1}{2} \left( g_{W^+ W^- Z} \, g_{W^+ \phi^{-}_i \phi^{0}_j} \right) = 0 \, ,
\end{align}
and for $a=W^+$, $b=Z'$, $i=\phi^{-}_i$, $j=\phi^{0}_j$:
\begin{align}
& \sum_k g_{W^+ \phi^{-}_i \phi^{0}_k} \, g_{Z' \phi^{0}_k \phi^{0}_j} - \frac{1}{2}
g_{W^+ Z' \phi^{-}_i \phi^{0}_j} + \frac{1}{4} \left( \frac{g_{W^+ Z \phi_i^{-}} \, 
g_{Z Z' \phi^{0}_j}}{m_Z^2} + \frac{g_{W^+ Z' \phi_i^{-}} \, 
g_{Z' Z' \phi^{0}_j}}{m_{Z'}^2} \right) \nonumber \\[1mm]
& + \frac{1}{2} \left( g_{W^+ W^- Z'} \, g_{W^+ \phi^{-}_i \phi^{0}_j} \right) = 0 \, ,
\end{align}
which concludes the list of the most relevant sum rules.
Note that because $g_{W^+ Z \phi_i^{-}} = 0$ in the SM and in many models beyond the SM,
most of the rules that include these couplings are sensitive to new physics.

\section{\label{sec:fermions}Bosons and fermions in SU(2)$_L \times $U(1)$_Y \times $U(1)$_{Y'}$}

For the purpose of simplicity, we will separate various interesting cases in the context of this model.

\subsection{\label{subsec:fermions_rule1}Rule 1}

In both ref.~\cite{Gunion:1990kf} and, with more detailed calculations,
in the appendix of \cite{Bento:2018fmy},
we see that the $s^1$ behavior of $\bar{f}_m f_n \rightarrow A_a A_b$ at large energies
is canceled through the equation
\begin{equation}\label{eq:6}
    \sum_p \left( g_{b \bar{m} p}^L \, g_{a \bar{p} n}^L - g_{a \bar{m} p}^L \, g_{b \bar{p} n}^L \right) 
    = \sum_e g_{a b e} \, g_{\bar{e} \bar{m} n}^L \, .
\end{equation}
Choosing $a=W^+$, $b=W^-$ and $m=n$ we have
\begin{equation}
    \sum_p \left( g_{W^- \bar{n} p}^L \, g_{W^+ \bar{p} n}^L - g_{W^+ \bar{n} p}^L \, g_{W^- \bar{p} n}^L \right) 
    = ( g_{W^+ W^- \gamma} \, g_{\gamma \bar{n} n}^L + g_{W^+ W^- Z} \, g_{Z \bar{n} n}^L 
    + g_{W^+ W^- Z'} \, g_{Z' \bar{n} n}^L ) \, ,
\end{equation}
where we may now consider one generation of fermions, as it simplifies the results.
Using $n=d$ and $p=u$ for quarks, we get
\begin{equation}
    \left( g_{W^- \bar{d} u}^L \, g_{W^+ \bar{u} d}^L - g_{W^+ \bar{d} u}^L \, g_{W^- \bar{u} d}^L \right) 
    = ( g_{W^+ W^- A} \, g_{A \bar{d} d}^L 
    + g_{W^+ W^- Z} \, g_{Z \bar{d} d}^L + g_{W^+ W^- Z'} \, g_{Z' \bar{d} d}^L ) \, .
\end{equation}
Choosing $n=\ell$ and $p=\nu$ for the charged lepton and its neutrino, we get
\begin{equation}
    \left( g_{W^- \bar{\ell} \nu}^L \, g_{W^+ \bar{\nu} \ell}^L 
    - g_{W^+ \bar{\ell} \nu}^L \, g_{W^- \bar{\nu} \ell}^L \right) 
    = ( g_{W^+ W^- A} \, g_{A \bar{\ell} \ell}^L 
    + g_{W^+ W^- Z} \, g_{Z \bar{\ell} \ell}^L + g_{W^+ W^- Z'} \, g_{Z' \bar{\ell} \ell}^L ) \, .
\end{equation}
Although many more sum rules can be derived [even prior to making use of eq.~\eqref{eq:4}],
if we choose $a=\bar{b}$ and $n=m$, one can
employ eq.~\eqref{eq:6}
in order to remove the triple gauge vertex
(cf. ref.~\cite{Gunion:1990kf}).
This also means that to extract the most useful information concerning $Z'$,
one must consider it as an external state,
which involves many couplings to the new gauge boson.
In particular, if we want to study quantitative or qualitative
properties of $Z'$,
we must relate its couplings mostly to SM interactions.
The other possibilities with $a\neq \bar{b}$ have the same shortcoming.
There is a triple gauge boson vertex but it will involve many unknowns.
Yet another possibility is to consider $a=W^-$ and $b=Z$,
but this rule does not contain information on $Z'$.

\subsection{\label{subsec:fermions_rule2}Rule 2}

For $m=n$, $a=Z$ and $i=\phi_i^0$:
\begin{align}
    & \frac{m_n}{2 m_Z^2} g_{Z Z \phi_i^0} ( g_{Z \bar{n} n}^R - g_{Z \bar{n} n}^L)
    + \frac{m_n}{2 m_{Z'}^2} g_{Z Z' \phi_i^0} ( g_{Z' \bar{n} n}^R - g_{Z' \bar{n} n}^L)
    \nonumber \\[1mm]
    & - \sum_k g_{Z \phi_i^0 \phi_k^0} \, g_{\phi_k^0 \bar{n} n}^L =
\sum_p 
( g_{\phi_i^0 \bar{n} p}^L \, g_{Z \bar{p} n}^L -
  g_{Z \bar{n} p}^R \, g_{\phi_i^0 \bar{p} n}^L) \,.
\end{align}
For simplicity, we again consider the case of one generation, where the sum over $p$ yields only one term $p=n$.
For example, if $n=f$ (where $f=u$ or $d$) then 
%
\begin{align}
    & \frac{m_f}{2 m_Z^2} g_{Z Z \phi_i^0} ( g_{Z \bar{f} f}^R - g_{Z \bar{f} f}^L)
    + \frac{m_f}{2 m_{Z'}^2} g_{Z Z' \phi_i^0} ( g_{Z' \bar{f} f}^R - g_{Z' \bar{f} f}^L)
    \nonumber \\[1mm]
    & - \sum_k g_{Z \phi_i^0 \phi_k^0} \, g_{\phi_k^0 \bar{f} f}^L =
    ( g_{\phi_i^0 \bar{f} f}^L \, g_{Z \bar{f} f}^L - g_{Z \bar{f} f}^R \, g_{\phi_i^0 \bar{f} f}^L) \, .
\end{align}
The case in which $a=Z'$ is less informative, as every coupling in the rule is dependent on the $Z'$.
It is straightforward to compute it, similarly as with $Z$.

For $m=n=u$, $a=W^+$ and $i=\phi^-_i$ we get
\begin{align}\label{eq:ferm1}
    & \frac{m_u}{2 m_Z^2} g_{W^+ Z \phi_i^-} ( g_{Z \bar{u} u}^R - g_{Z \bar{u} u}^L)
    + \frac{m_u}{2 m_{Z'}^2} g_{W^+ Z' \phi_i^-} ( g_{Z' \bar{u} u}^R - g_{Z' \bar{u} u}^L)
    \nonumber \\[1mm]
    & = \sum_k g_{W^+ \phi^-_i \phi_k^0} \, g_{\phi_k^0 \bar{u} u}^L \, ,
\end{align}
and, exchanging $L\leftrightarrow R$, we find
\begin{align}
    & \frac{m_u}{2 m_Z^2} g_{W^+ Z \phi_i^-} ( g_{Z \bar{u} u}^L - g_{Z \bar{u} u}^R)
    + \frac{m_u}{2 m_{Z'}^2} g_{W^+ Z' \phi_i^-} ( g_{Z' \bar{u} u}^L - g_{Z' \bar{u} u}^R)
    \nonumber \\[1mm]
    & = \sum_k g_{W^+ \phi^-_i \phi_k^0} \, g_{\phi_k^0 \bar{u} u}^L - g_{W^+ \bar{u} d}^L \, g_{\phi_i^- \bar{d} u} \, .
\end{align}
For $m=n=d$, $a=W^+$ and $i=\phi^-_i$ we get
\begin{align}
    & \frac{m_d}{2 m_Z^2} g_{W^+ Z \phi_i^-} ( g_{Z \bar{d} d}^R - g_{Z \bar{d} d}^L)
    + \frac{m_d}{2 m_{Z'}^2} g_{W^+ Z' \phi_i^-} ( g_{Z' \bar{d} d}^R - g_{Z' \bar{d} d}^L)
    \nonumber \\[1mm]
    & = \sum_k g_{W^+ \phi^-_i \phi_k^0} \, g_{\phi_k^0 \bar{d} d}^L + g_{\phi_i^- \bar{d} u}^L
    \, g_{W^+ \bar{u} d}^L \, ,
\end{align}
and, exchanging $L\leftrightarrow R$, we find
\begin{align}
    & \frac{m_d}{2 m_Z^2} g_{W^+ Z \phi_i^-} ( g_{Z \bar{d} d}^L - g_{Z \bar{d} d}^R)
    + \frac{m_d}{2 m_{Z'}^2} g_{W^+ Z' \phi_i^-} ( g_{Z' \bar{d} d}^L - g_{Z' \bar{d} d}^R  )
    \nonumber \\[1mm]
    & = \sum_k g_{W^+ \phi^-_i \phi_k^0} \, g_{\phi_k^0 \bar{d} d}^R  \, .
\end{align}
This concludes all useful sum rules with fermions.

\section{\label{sec:generic}Generic applications}

Some of the most interesting applications of tree-level unitarity are
the ones that need the least information or where the information is better known.
We begin with the rule of eq.~\eqref{eq:sum1}.
By assuming a theory of scalar singlets and doublets,
we may already drop any coupling with $\phi_i^{\pm \pm}$.
In these models, eq.~\eqref{eq:sum1} takes the form
\begin{equation} \label{WWArule}
    4m_W^2 \, g_{W^+ W^- \gamma}^2 + (4m_W^2 - 3m_Z^2) g_{W^+ W^- Z}^2 +
    (4m_W^2 -3m_{Z'}^2) g_{W^+ W^- Z'}^2 = \sum_k g_{W^+ W^- \phi^0_k}^2 \, ,
\end{equation}
where the third term in the left-hand side of \eq{WWArule} differentiates this sum rule from the one
in the SM.  Generalizing the electroweak $\rho$ parameter of the SM,
we may use \eq{WWArule} to define a new parameter $\rho'$. We will
discuss the importance and definition of this parameter in 
section~\ref{sec:a_new_rho}.

Eq.~\eqref{eq:sum5} provides another interesting sum rule.
We can provide a qualitative interpretation of this sum rule as follows.
Let us define the Higgs boson as $\phi^0_1$.
Then, if $g_{Z Z \phi_k^0} \sim 0$, for $k>1$,
and $g_{Z' Z' \phi_1^0} \sim 0$
(as one might expect from a hidden sector
type of model),
we conclude that 
\begin{equation}
\sum_k g_{Z Z' \phi^0_k}^2 \sim 0 \, .
\end{equation}
Although this is not a very strong statement, it does give a way of
probing aspects of hidden sectors through $g_{Z Z' \phi^0_k}$.

From eqs.~\eqref{eq:sum11} and
\eqref{eq:sum12}, by assuming couplings of the type $WZ\phi$ and $WZ'\phi$ to be zero, 
or at least approximately zero, we extract that
\begin{eqnarray}
    \sum_k g_{W^- \phi_i^+ \phi^0_k} g_{Z Z \phi^0_k} &=& 0 \, ,
\nonumber\\*[1mm]
    \sum_k g_{W^- \phi_i^+ \phi^0_k} g_{Z' Z' \phi^0_k} &=& 0 \, .
\label{ZZphi0}
\end{eqnarray}
This property is analogous to the one obtained for the $Z$ boson in a NHDM
with a SU(2)$_L \times $U(1)$_Y$ gauge group \cite{Bento:2017eti}.

If the $WZ\phi$ and $WZ'\phi$ vertices are absent, then the sum rule
given in eq.~\eqref{eq:ferm1} yields
\begin{equation}
    \sum_k g_{W^+ \phi_i^- \phi^0_k} \, g_{\phi_k^0 \bar{u} u}^L = 0 \, .
\end{equation}
Comparing with eqs.~\eqref{ZZphi0},
suggests a connection between the couplings
$g_{Z Z \phi^0_k}$,
$g_{Z' Z' \phi^0_k}$ and $g_{\phi_k^0 \bar{u} u}^L$,
which are all orthogonal to $W \phi \phi$ type couplings.

The sum rules of this section apply to any $Z^\prime$ model. 
In specific cases where expansions are performed, the sum rules can be used in
order to check for the consistency of the corresponding expansions,
as illustrated in eq.~\eqref{eq:wrong_unitarity} below.

\section{\label{sec:models}Sum rules as model consistency checks}

In this section,
we study one particular sum rule in the context of
a model for a dark sector that is common in the literature~\cite{Curtin:2014cca,Babu:1997st,Davoudiasl:2014mqa,Davoudiasl:2012ag,Cheng:2022aau,Foguel:2022unm}.
We shall examine a model with an electroweak gauge group SU(2)$_L\times$U(1)$_Y\times$U(1)$_{Y'}$, 
where all SM particles are neutral under
the U(1)$_{Y'}$ [which is often referred to in the literature as U(1)$_D$], corresponding to the dark sector of the model.
The gauge boson associated with U(1)$_D$ will henceforth be denoted by $Z_D$.
In addition, we add to the model one extra scalar singlet
$S$ that is charged under U(1)$_D$, with $Y' = 1$.
In this model, there are no gauge anomalies.

We shall assume that, when the scalar potential of the model
is minimized, the singlet field $S$ acquires a vacuum expectation value,
\begin{equation}
\vev{S}=\frac{v_D}{\sqrt{2}}\, .
\end{equation}
We then define the following dimensionless ratio,
\beq \label{deltadef}
\delta\equiv \frac{2g_D v_D}{(g^2+g^{\prime\,2})^{1/2}v}\,,
\eeq
where $v\simeq 246$~GeV is the vacuum expectation value of the SM Higgs field
and $g$, $g'$ are the SU(2)$_L$ and U(1)$_Y$ gauge couplings of the 
SM electroweak Lagrangian, respectively.

After introducing the singlet scalar field $\hat S^0$ via
\beq \label{shatdef}
 S =\frac{1}{\sqrt{2}}\left(v_D + \hat{S}^0\right)\,,
 \eeq
we note that $\hat S^0$ can mix with the would-be physical Higgs boson of the SM (denoted by~$\phi^0$).  The physical scalar mass eigenstates $h$ and $S^0$ are given by
\begin{equation}
    \begin{pmatrix}
    h \\
    S^0
    \end{pmatrix}
    =
    \begin{pmatrix}
    c_h & \,\,\, -s_h \\
    s_h &\,\,\, \phantom{-}c_h
    \end{pmatrix}
    \begin{pmatrix}
    \phi^0 \\
    \hat{S}^0
    \end{pmatrix} \,,
\label{hS_mixing}
\end{equation}
where $\theta_h$ is the corresponding mixing angle, 
$c_h\equiv \sin\theta_h$, and $s_h\equiv\sin\theta_h$.

\subsection{The gauge sector Lagrangian}
\label{gaugesector}

We begin with the Lagrangian
\begin{equation}
    \mathcal{L} \supset 
     -\frac{1}{4} W^a_{\mu \nu} W^{a \mu \nu}    
    - \frac{1}{4} \hat{B}_{\mu \nu} \hat{B}^{\mu \nu}
    - \frac{1}{4} \hat{X}_{\mu \nu}  \hat{X}^{\mu \nu} +
\frac{\epsilon}{2 c_W} \hat{X}_{\mu \nu} \hat{B}^{\mu \nu} \, ,
\label{hatBhatX_to_BX}
\end{equation}
where the SU(2)$_L$ gauge field strength tensor is given by \pagebreak
\begin{equation}
W^a_{\mu \nu} = \partial_\mu W^a_\nu - \partial_\nu W^a_\mu - g \, 
\epsilon^{abc} \, W^b_\mu \,
W^c_\nu \, ,
\end{equation}
and the U(1)$_{Y}$ and U(1)$_{Y'}$ gauge field strength tensors are respectively given by
\beq
\hat{B}_{\mu\nu}= \partial_\mu \hat{B}_\nu - \partial_\nu \hat{B}_\mu\,,
\qquad\quad \hat{X}_{\mu\nu}= \partial_\mu \hat{X}_\nu - \partial_\nu \hat{X}_\mu\,.
\eeq
The Lagrangian exhibited in \eq{hatBhatX_to_BX} includes a kinetic mixing term that is governed by a parameter
$\epsilon$.  Phenomenological considerations suggest that $|\epsilon|\ll 1$.\footnote{For a compilation of the most
recent bounds on $\epsilon$, see ref.~\cite{ParticleDataGroup:2022pth}.}
At the pure gauge level, there is no distinction between the fields $\hat{B}$ and $\hat{X}$.
What will distinguish between $\hat{B}$ and $\hat{X}$ will be
their differing couplings to fermion and scalar fields.

One can obtain canonical kinetic gauge terms by making the transformation
\begin{align}
    &\hat{X}^\mu = \eta X^\mu \, , \nonumber \\
    &\hat{B}^\mu = B^\mu + \frac{\epsilon}{c_W} \eta X^\mu \, ,
\end{align}
where
\begin{equation} \label{etadef}
\eta\equiv \frac{1}{\sqrt{1 - \epsilon^2/c_W^2}}\,.
\end{equation}
At this stage, $c_W$ is just a convenient notation with no physical meaning.
The parameter $c_W$ will acquire physical meaning in eq.~\eqref{W3B_to_AZ} below.

After these field redefinitions, we may rotate the gauge fields to get
the SM photon and a SM-like field $Z^0_\mu$.
We start from the covariant derivative
\begin{equation}
    D_\mu = \partial_\mu + i g \left(T^+ W^+_\mu + T^- W^-_\mu \right) + i g T_3 W^3_\mu
    + i g' Y \hat{B}_\mu + i g_D Y' \hat{X}_{\mu} \,,
\end{equation}
where $W_\mu^\pm=(W_\mu^1\mp iW_\mu^2)/\sqrt{2}$.
After employing \eqref{hatBhatX_to_BX},
\begin{equation}\label{eq:ew_rot}
    D_\mu = \partial_\mu + i g \left(T^+ W^+_\mu + T^- W^-_\mu \right) + i g T_3 W^3_\mu
    + i g' Y B_\mu 
    + i \left( g' \frac{\epsilon}{c_W} \eta Y + g_D Y' \eta \right) X_\mu \, .
\end{equation}
When acting on SU(2)$_L$ doublet fields,
\begin{equation}
T^+
= \frac{1}{\sqrt{2}}
\left(
\begin{array}{cc}
0 &\  \phm  1\\
0 &\   \phm 0
\end{array}
\right)\, ,
\ \ \ \ 
T^-
= \frac{1}{\sqrt{2}}
\left(
\begin{array}{cc}
0 &\   \phm 0\\
1 &\  \phm  0
\end{array}
\right)\, ,
\ \ \ \ 
T_3
= \frac{1}{2}
\left(
\begin{array}{cc}
1 &\  \phm 0\\
0 &\  -1
\end{array}
\right)\, .
\end{equation}

By introducing a scalar doublet $\Phi^T = \left(\phi^+, (v + \phi^0)/\sqrt{2}\right)$,
one can diagonalize the quadratic terms of the Lagrangian
\begin{equation}
    |D_\mu \Phi|^2 \supset 
\left(\frac{gv}{2}\right)^2 W^+_\mu W^{-\, \mu}
+
\frac{1}{8} v^2 \left[ g^2 \left( W^3_\mu \right)^2
    - 2 g g' W_\mu^3 B^\mu + g'^2 \left( B_\mu \right)^2 \right] \, .
\end{equation}
Thus, the $W$ gauge boson coincides with the SM one, with mass
\begin{equation}
m_W=\frac12 gv\, .
\end{equation}
Next, we define rotated fields
\begin{align}
    &Z_\mu^0 = c_W\, W^3_\mu  - s_W\, B_\mu  \, , \nonumber \\
    &A_\mu = s_W\, W^3_\mu + c_W\, B_\mu \, ,
\label{W3B_to_AZ}
\end{align}
where $s_W\equiv\sin\theta_W$ and $c_W\equiv\cos\theta_W$,
which defines $g'=g t_W$  (with $t_W\equiv s_W/c_W$) and the angle $\theta_W$
as the angle that rotates to a basis where there is a massless gauge field $A_\mu$ (to be identified with the photon $\gamma$).
Then, the $Z^0$ field has the couplings of the SM massive neutral gauge boson,
and we define
\begin{equation}
m_{Z^0} = \frac{g v }{2 c_W} = \frac{m_W}{c_W}.
\label{mZ_def}
\end{equation}
We emphasize that the interaction eigenstate field $Z^0$ does \textit{not} correspond to the experimentally observed $Z$ gauge boson since it is \textit{not}
a mass eigenstate field. However, note that the couplings of $Z^0$
coincide with those of
the massive neutral gauge boson of the Standard Model.
Finally, we get for the covariant derivative
\begin{align}\label{eq:new_u1_rot}
    D_\mu =& \partial_\mu + i g \left(T^+ W^+_\mu + T^- W^-_\mu \right) + i e Q A_\mu
    + i \frac{g}{c_W} \left(T_3 - Q s_W^2\right) Z^0_\mu \nonumber \\
    &+ i \left( g t_W \frac{\epsilon}{c_W} \eta Y + g_D Y' \eta \right) X_\mu \, ,
\end{align}
while keeping in mind that $Z^0$ and $X$ are interaction eigenstate fields
that must eventually be re-expressed in terms of mass eigenstate vector boson fields.

The remaining scalar kinetic terms are simple to obtain. We are interested
in the mass terms of the remaining massive neutral gauge bosons. 
The covariant derivative acts on the scalars
according to their charge, such that
\begin{align}
    &D_\mu \Phi = \left[\partial_\mu + \cdots + i \frac{g}{c_W} T_3 Z^0_\mu 
    + i g t_W \frac{\epsilon}{2c_W} \eta X_\mu \right] \Phi \, , \nonumber \\
    &D_\mu S = \left[ \partial_\mu
    + i g_D \eta X_\mu \right] S \, ,
\end{align}
where the scalar field $S$ is defined in \eq{shatdef}.
Then,
\begin{equation}
    |D_\mu \Phi|^2 \supset
    m_{Z^0}^2
    \left[ \frac{1}{2} (Z^0_\mu)^2
    - (\eta t_W \epsilon) Z^0_\mu X^\mu 
    + \frac{1}{2} (\eta t_W \epsilon)^2 (X^\mu)^2 \right] \, ,
\end{equation}
and
\begin{equation}
    |D_\mu S|^2 \supset \frac{1}{2} g_D^2 v_D^2 \eta^2 (X^\mu)^2
    = \frac{1}{2} m_{Z^0}^2 \eta^2 \delta^2 (X^\mu)^2\, ,
\end{equation}
where $\delta = g_D v_D / m_{Z^0}$ in light of \eqs{deltadef}{mZ_def}.
With these definitions, we obtain the squared mass matrix 
of the neutral massive gauge bosons:\footnote{Note that in ref.~\cite{Foguel:2022unm},
the $22$ element of the squared mass matrix $\mathcal{M}^2_{Z Z_D}$ is incorrectly given by
 \mbox{$m_{Z^0}^2(\eta^2 t^2_W \epsilon^2 + \delta^2)$,} where definitions of $m_{Z^0}^2$, $t_W$, $\delta$, and $\eta$ are the same as the ones employed in
 this paper (and $\epsilon$ is defined with an opposite sign).  To obtain the correct expression, one must replace $\delta^2$ with $\eta^2\delta^2$.  A similar
 correction must also be applied to the expression for $\tan 2\alpha$ given in ref.~\cite{Foguel:2022unm}, where $\alpha$ is the mixing angle defined
 in \eqs{zzerodef}{Xdef}.
\label{foot3}}
\begin{equation} \label{sqmassmat}
    \mathcal{M}^2_{Z Z_D} \equiv\begin{pmatrix} m_{Z^0}^2 & \,\,\, m_{XZ}^2 \\[3pt]
    m_{XZ}^2 & \,\,\, m_X^2\end{pmatrix}= m_{Z^0}^2
    \begin{pmatrix}
    1 &\quad -\eta t_W \epsilon \\[3pt]
    -\eta t_W \epsilon &\quad  \eta^2 t^2_W \epsilon^2 + \eta^2 \delta^2
    \end{pmatrix} \, .
\end{equation}
One can now use an orthogonal matrix to diagonalize
the mass matrix such that
\begin{align}
    &Z^0 = Z \cos \alpha 
    - Z_D \sin \alpha\, , \label{zzerodef} \\
    &X = Z \sin \alpha
    + Z_D \cos \alpha \, ,\label{Xdef}
\end{align}
where $Z$ and $Z_D$ are mass eigenstates with squared masses,
\begin{align}
    &m_Z^2 = m_{Z^0}^2 \left[ 1 - \sin^2 \alpha \left(1 - \delta ^2 \eta^2\right)
    +\eta t_W \epsilon  \sin \alpha \left(\eta t_W \epsilon  \sin \alpha  -2 \cos \alpha  
    \right) \right] \,, \label{mzee} \\[3pt]
    &m_{Z_D}^2 = m_{Z^0}^2 \left[ \sin ^2\alpha \left(1-\delta ^2 \eta ^2\right)
    +\delta ^2 \eta ^2+\eta t_W \epsilon  \cos \alpha  \left(2 \sin \alpha
    +\eta t_W \epsilon  \cos \alpha \right) \right] \,,\label{mzeedee}
\end{align}
where the mixing angle $\alpha$ can be chosen to lie in the range $-\tfrac12\pi<\alpha\leq\tfrac12\pi$, with
\begin{align} 
\sin 2\alpha&=\frac{- 2 \xi\eta t_W \epsilon}{\sqrt{\left[1 - (\eta t_W \epsilon)^2 - \eta^2 \delta^2\right]^2 + 4 (\eta t_W \epsilon)^2}}\,,\label{sintwoa}
\\[4pt]
\cos 2 \alpha& = \frac{\xi[1 - (\eta t_W \epsilon)^2 - \eta^2 \delta^2]}
    {\sqrt{\left[1 - (\eta t_W \epsilon)^2 - \eta^2 \delta^2\right]^2 + 4 (\eta t_W \epsilon)^2}}
    \,,\label{costwoa}
\end{align}
and
\beq \label{xidef}
\xi\equiv \sgn(m_Z^2-m_{Z_D}^2)=\begin{cases} +1 & \text{if $m_Z>m_{Z_D}$}\,,\\
-1 & \text{if $m_Z<m_{Z_D}$}\,.\end{cases}
\eeq
Using \eqs{sintwoa}{costwoa}, one can then derive
\begin{align}
\cos\alpha & =\left( \frac{\xi[1-(\eta t_W \epsilon)^2 - \eta^2 \delta^2]+\sqrt{\left[1 - (\eta t_W \epsilon)^2 - \eta^2 \delta^2\right]^2 + 4 (\eta t_W \epsilon)^2}}{2\sqrt{\left[1 - (\eta t_W \epsilon)^2 - \eta^2 \delta^2\right]^2 + 4 (\eta t_W \epsilon)^2}}\right)^{1/2}\,,\\[4pt]
\sin\alpha & =  {\rm sgn}(-\xi\epsilon)\sqrt{1-\cos^2\alpha}\,.
\label{aux22}
\end{align}
Note that, in light of \eqst{mzee}{costwoa}, it follows that
\beq \label{mzdiff}
m_Z^2-m_{Z_D}^2=\xi m_{Z^0}^2\sqrt{\left[1 - (\eta t_W \epsilon)^2 - \eta^2 \delta^2\right]^2
+ 4 (\eta t_W \epsilon)^2}\,,
\eeq
which is consistent with the definition of $\xi$ given in \eq{xidef}.

One can also derive expressions for the elements of $\mathcal{M}^2_{ZZ_D}$ in terms of the physical parameters $m_Z^2$, $m_{Z_D}^2$ and $\alpha$:
\beqa
    m_{Z^0}^2 &=& m_Z^2 \cos^2 \alpha + m_{Z_D}^2 \sin^2 \alpha\,,\label{mzzero}\\
   m_{X}^2 &=& m_Z^2 \sin^2 \alpha + m_{Z_D}^2 \cos^2 \alpha \,,\label{mxzero}\\
   m_{XZ}^2 &=& \half(m_Z^2-m_{Z_D}^2)\sin 2\alpha\,.\label{mxz}
\eeqa
In terms of the physical fields,
the covariant derivative in eq.~\eqref{eq:new_u1_rot} becomes,
\begin{align}\label{eq:new_u1_rot2}
    D_\mu =&\ \partial_\mu + i g \left(T^+ W^+_\mu + T^- W^-_\mu \right) + i e Q A_\mu
    \nonumber \\
    &+ \left[ i \frac{g}{c_W} \left(T_3 - Q s_W^2\right) c_\alpha
    + i \left( g t_W \frac{\epsilon}{c_W} \eta Y + g_D Y' \eta \right) s_\alpha \right]  Z_\mu 
    \nonumber \\
    &+ \left[  i \left( g t_W \frac{\epsilon}{c_W} \eta Y + g_D Y' \eta \right) c_\alpha
    - i \frac{g}{c_W} \left(T_3 - Q s_W^2\right) s_\alpha
     \right] Z_{D \mu} \, .
\end{align}

For the convenience of the reader, we list all the relevant model parameters in table~\ref{parms}.
The relations between our notation
and the notation employed in
ref.~\cite{Babu:1997st} are provided in appendix~\ref{app:notations}.

\begin{table}[t!]
\centering
\begin{tabular}{|c|c|}\hline
parameter &definition \\ \hline
$v$ & $246~{\rm GeV}$ \\[3pt]
$v_D$ & $\sqrt{2}\langle S\rangle$ \\[3pt]
$g$, $g'$, $g_D$ & SU(2)$_L\times$U(1)$_Y\times$U(1)$_{Y'}$ gauge couplings \\[3pt]
$\epsilon$ & gauge kinetic mixing parameter \\[3pt]
$m_{Z^0}$ & $\half (g^2+ g^{\prime\,2})^{1/2} v$\\[3pt]
$\delta$ & $g_D v_D/m_{Z^0}$ \\[3pt]
$m_Z$ & mass of the physical (observed) $Z$ boson \\[3pt]
$m_{Z_D}$ & mass of the physical dark $Z$ boson \\[3pt]
$\xi$ & $\sgn(m_Z^2-m_{Z_D}^2)$ \\[3pt]
$c_W$ & $m_W/m_{Z^0}$ \\[3pt]
$t_W$ & $(1-c_W^2)^{1/2}/c_W$ \\[3pt]
$\eta$ & $1/(1-\epsilon^2/c_W^2)^{1/2}$ \\[3pt]
$m_{X}^2$ & $m_{Z^0}^2\bigl[(\eta t_W\epsilon)^2+\eta^2\delta^2\bigr]$ \\[3pt]
$m_{XZ}^2$ & $-\xi\eta\epsilon t_W m_{Z^0}^2$ \\[3pt]
$\alpha$ & $Z^0$--$X$ mixing angle \\[3pt]
$c_h$, $s_h$ & cosine and sine of the $\phi^0$--$\hat{S}^0$ mixing angle \\
\hline
\end{tabular}
\caption{\small A list of the parameters that govern 
the SU(2)$_L\times$U(1)$_Y\times$U(1)$_{Y'}$ model. 
\label{parms}}
\end{table}

\subsubsection{Expansion in $\epsilon$}

Starting with the squared mass matrix given in \eq{sqmassmat}, we can expand in $\epsilon$ to obtain
\begin{equation}
    \mathcal{M}^2_{Z Z_D} = m_{Z^0}^2
    \begin{pmatrix}
    1 &\ \  -t_W \epsilon \\[10pt]
    -t_W \epsilon &\ \  \delta^2 + \epsilon^2 \left( \frac{\delta^2}{c_W^2} + t_W^2 \right)
    \end{pmatrix} + \mathcal{O}\left( \epsilon^3 \right) \,.
\end{equation}
Under the assumption that $|\epsilon|\ll 1$, one can approximate
\beq
\xi=\sgn(1-\delta^2)\,,
\eeq
where $\xi$ is defined in \eq{xidef}.
Moreover, one
must assume that $1-\delta^2$ is not too small.  Otherwise, the
two eigenvalues of $\mathcal{M}^2_{Z Z_D}$ would be nearly degenerate, and the perturbative analysis that follows would be invalid.

Under the assumptions that $|\epsilon|\ll 1$ and  $1-\delta^2\sim\mathcal{O}(1)$,
the squared masses of the physical gauge bosons given in \eqs{mzee}{mzeedee} yield
\begin{align}
    &m_Z^2 = m_{Z^0}^2 \left[ 1 + \frac{\epsilon^2 t_W^2}{1-\delta^2}
    + \mathcal{O}\left( \epsilon^4 \right) \right] \, , \label{eq:masses_exp1}\\[6pt]
    &m_{Z_D}^2 = m_{Z^0}^2 \left[ 1 + \epsilon^2 \left(
    \frac{1}{c_W^2} - \frac{t_W^2}{1-\delta^2}\right)
    + \mathcal{O}\left( \epsilon^4 \right) \right] \delta^2  \,.\label{eq:masses_exp2}
\end{align}
Likewise, the interaction eigenstate fields $Z^0$ and $X$ given in \eqs{zzerodef}{Xdef} can be expressed in terms of the mass eigenstate
fields,
\begin{align}
    &Z^0 = \left[ 1 - \frac{\epsilon^2 t_W^2}{2(1-\delta^2)^2}
    + \mathcal{O}\left(\epsilon^4\right) \right] Z
    + \left[ \frac{\xi\epsilon t_W}{1 - \delta^2} 
    + \mathcal{O}\left(\epsilon^3\right) \right] Z_D \, ,  \label{Zmasseigen}\\[6pt]
    &X = \left[ - \frac{\xi\epsilon t_W}{1 - \delta^2} 
    + \mathcal{O}\left(\epsilon^3\right) \right] Z
    + \left[ 1 - \frac{\epsilon^2 t_W^2}{2(1-\delta^2)^2}
    + \mathcal{O}\left(\epsilon^4\right) \right] Z_D \, .\label{Xmasseigen}
\end{align}

\subsection{Unitarity sum rule}\label{sec:unitarity_sum_rule}

To check the consistency of the model, we shall verify that the sum rule obtained in \eq{WWArule}
as a consequence of tree-level unitarity is satisfied.
For the convenience of the reader, we repeat here
the sum rule given by \eq{WWArule}: 
\begin{equation}\label{eq:sum_rule}
    4 m_W^2 \, g_{W^+ W^- \gamma}^2 + (4m_W^2 - 3m_Z^2) \, g_{W^+ W^- Z}^2
    + (4m_W^2 - 3m_{Z_D}^2) \, g_{W^+ W^- Z_D}^2 = \sum_k g_{W^+ W^- \phi^0_k}^2
    \,.
\end{equation}
Eq.~\eqref{eq:sum_rule} must be satisfied both exactly and order by order in
the mixing parameter~$\epsilon$, which will serve as a good check of our computations.

\subsubsection{Exact sum rule}

The sum rule exhibited in \eq{eq:sum_rule} follows from the substitution of the
parameters
\begin{align}
    &g_{W^+ W^- \gamma}^2 = g^2 s_W^2 \, ,\label{eq:def_parameters_sr1} \\
    &g_{W^+ W^- Z}^2 = c_\alpha^2 g^2 c_W^2 \, ,\label{eq:def_parameters_sr2} \\
    &g_{W^+ W^- Z_D}^2 = s_\alpha^2 g^2 c_W^2 \, , \label{eq:def_parameters_sr3} \\
    &g_{W^+ W^- h}^2 = g^2 m_W^2 c_h^2 \, , \label{eq:def_parameters_sr4}\\
    &g_{W^+ W^- S^0}^2 = g^2 m_W^2 s_h^2 \, ,\label{eq:def_parameters_sr5}
\end{align}
into eq.~\eqref{eq:sum_rule}, where $c_\alpha\equiv\cos\alpha$ and $s_\alpha\equiv\sin\alpha$.  The end result is
\begin{equation}\label{eq:first_sum_rule_cs}
    4 m_W^2 - 3 c_W^2 \left( m_Z^2 \cos^2 \alpha + m_{Z_D}^2 \sin^2 \alpha \right) = m_W^2
    \, ,
\end{equation}
or equivalently
\begin{equation}\label{eq:second_sum_rule_cs}
    4 m_W^2 - \frac{3}{2} c_W^2 \bigl[ m_Z^2 + m_{Z_D}^2 
    + (m_Z^2 - m_{Z_D}^2) \cos 2 \alpha \bigr] = m_W^2
    \, .
\end{equation}
Using eqs.~\eqref{mzee}--\eqref{aux22} in \eq{eq:first_sum_rule_cs},
we find
\begin{equation}
    4 m_W^2 - \frac{3}{2} c_W^2 m_{Z^0}^2 
    \left[ \left(1 + (\eta t_W \epsilon)^2 + \eta^2 \delta^2 \right)
    + \left(1 - (\eta t_W \epsilon)^2 + \eta^2 \delta^2 \right) \right] = m_W^2
    \, ,
\end{equation}
which simplifies to
\begin{equation}
    m_W^2 - c_W^2 m_{Z^0}^2 = 0
    \,.
\label{MZMWcW}
\end{equation}
Of course, \eq{MZMWcW} is true in light of eq.~\eqref{mZ_def}.

Alternatively, we can
work out eq.~\eqref{eq:first_sum_rule_cs} by using \eq{mzzero},
which again reproduces the result given in eq.~\eqref{MZMWcW}.

\subsubsection{Order by order sum rule in powers of $\epsilon$}

We now return to \eqref{eq:sum_rule}.
We expand the masses given in \eqs{eq:masses_exp1}{eq:masses_exp2}
and the mixing parameters to order $\mathcal{O}(\epsilon^2)$.
We substitute these
in the couplings of \eqst{eq:def_parameters_sr1}{eq:def_parameters_sr5},
to find
\begin{align}
    &g_{W^+ W^- Z} = g c_W \left( 1 - \frac{\epsilon^2 t_W^2}{2(1-\delta^2)^2} \right) 
    \, , \label{eq:VVV_expanded1} \\
    &g_{W^+ W^- Z_D} = g c_W \left( \frac{\xi\epsilon t_W}{1 - \delta^2} \right)  \,.\label{eq:VVV_expanded2}
\end{align}
We thus confirm the sum rule in \eqref{eq:sum_rule} to order $\mathcal{O}(\epsilon^2)$.
As explained in section~\ref{testing},  we have detected some cases in the literature in which
the sum rule \eqref{eq:sum_rule} fails due to the use of inconsistent approximations for the relevant couplings and/or masses.

\subsubsection{Defining the weak angle}

The weak mixing angle defined in \eq{W3B_to_AZ} is not equivalent to the corresponding quantity defined in an  SU(2)$_L\times$U(1)$_Y$ electroweak theory, since $Z^0$ is not a mass eigenstate.  
Since the Fermi constant and the fine structure constant are, respectively,
\begin{equation} \label{Fermi}
\frac{G_F}{\sqrt{2}}=\frac{g^2}{8m_W^2}\,,\qquad\quad \alpha\lsub{\rm EM}\equiv\frac{e^2}{4\pi}=\frac{g^2 s_W^2}{4\pi}\,,
\end{equation}
and are more precisely measured than $m_W$ and $g$, it is common practice to define $\theta_W$ via
\begin{equation}
    s_W^2 c_W^2 = \frac{\pi \alpha\lsub{\rm EM}} {\sqrt{2} G_F m_Z^2}\,.
\end{equation}
Clearly, such an option is not available in an SU(2)$_L\times$U(1)$_Y\times$U(1)$_{Y'}$ electroweak theory.  

However, to facilitate a comparison between the gauge
groups  SU(2)$_L\times$U(1)$_Y$ and SU(2)$_L\times$U(1)$_Y\times$U(1)$_{Y'}$,
one can instead define $\theta_W$ in terms of the following three input parameters
that are common to both theories:
the Fermi constant $G_F$,
the fine structure constant $\alpha\lsub{\rm EM}$,
and the $W$ boson mass $m_W$.  
Although $m_W$ is not as precisely measured as $m_Z$, the choice of $m_W$ is convenient since it does not preclude the possibility of additional U(1) gauge groups that weakly mix with the hypercharge U(1)$_Y$.
 At tree-level, \eq{Fermi} yields,
\begin{equation}\label{eq:definition_sw}
s^2_W=\frac{\pi\alpha\lsub{\rm EM}}{\sqrt{2} G_F m_W^2}\,,
\end{equation}
which then can be taken as an all-orders definition of $\theta_W$. 

Next, consider the definition of the $\rho$-parameter,
\beq \label{oldrho}
\rho\equiv\frac{m_W^2}{m_Z^2 c_W^2}\,.
\eeq
The dependence of $\rho$ on the weak mixing angle is inconvenient if we wish to use this parameter in both the SU(2)$_L\times$U(1)$_Y$ and SU(2)$_L\times$U(1)$_Y\times$U(1)$_{Y'}$ models.   In light of the discussion above, it is convenient to employ \eq{eq:definition_sw} to obtain a definition of $\rho$ that is suitable in both models,
\begin{equation}
    \rho \equiv \frac{2 G_F m_W^4}{m_Z^2 \left( 2 G_F m_W^2 - \sqrt{2} \pi \alpha\lsub{\rm EM}  \right)} \, .
\end{equation}
Indeed, the SM relation, $\rho=1$ will no longer hold in the 
SU(2)$_L\times$U(1)$_Y\times$U(1)$_{Y'}$, model, since the relation 
between $m_W$ and $m_Z$ is modified as compared to the Standard Model.
It is convenient to introduce a related parameter, denoted by $\rho^\prime$, whose tree-level value will be equal to 1
in the SU(2)$_L\times$U(1)$_Y\times$U(1)$_{Y'}$, model under the assumption
that the Higgs sector consists of SU(2)$_L$ doublets with $Y=\frac12$.

\subsubsection{A new $\rho'$ parameter}\label{sec:a_new_rho}

In analogy with \eq{oldrho}, we define $\rho^\prime$ in the SU(2)$\times$U(1)$\times$U(1)$^\prime$ model to be 1 at tree level as follows:
\begin{equation} \label{rhoprime}
    \rho' = \frac{m_W^2}{\left( m_Z^2 \cos^2 \alpha 
    + m_{Z_D}^2 \sin^2 \alpha \right) c_W^2} = 1 \,.
\end{equation}
Indeed, the statement that $\rho^\prime=1$ is simply a consequence of the
sum rule given in eq.~\eqref{eq:sum_rule}. In turn, eq.~\eqref{rhoprime} applies to 
a class of models where for the new multiplet of weak isospin $T$ and hypercharge $Y$ (with arbitrary $Y'$)
satisfies the equation $T(T + 1) = 3 Y^2$. 
Further details can be found in appendix~\ref{app:p_legs}.
One can rewrite \eq{rhoprime} in terms of the $\rho$-parameter defined in \eq{oldrho},
\begin{equation} \label{rhoone}
    \rho' = \frac{\rho}{\cos^2 \alpha 
    + \left(\frac{m_{Z_D}}{m_Z}\right)^2 \sin^2 \alpha }=1 \,.
\end{equation}
It then follows that
\cite{Altarelli:1991zg}
\begin{equation}
    \rho - 1 =
    \left[  \left( \frac{m_{Z_D}}{m_Z}\right)^2-1 \right]
    \sin^2 \alpha \, .
\label{master3}
\end{equation}
\begin{figure}[t!]
  \centering
\includegraphics[width=0.75\textwidth]{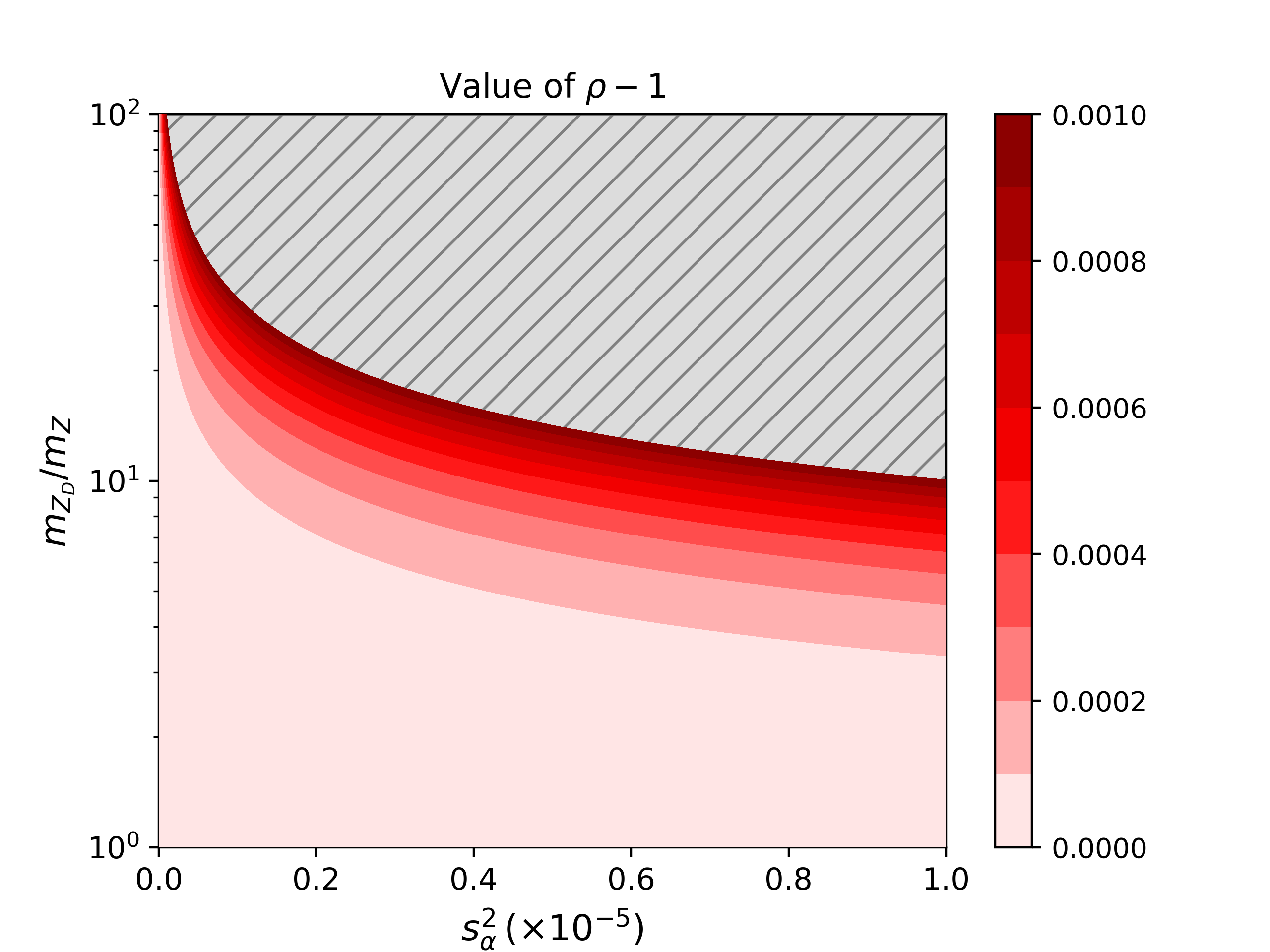}
\caption{Impact of \eq{master3} on the $(s_\alpha^2,\, m_{Z_D}/m_Z)$ plane,
for $\rho-1 \in [0, 1] \times 10^{-3}$. The gray area corresponds to $\rho - 1 > 10^{-3}$}
\label{fig:1}
\end{figure}

For a given value of $\rho - 1$,
\eq{master3} defines a line in the
$(s_\alpha^2,\, m_{Z_D}/m_Z)$ plane,
where $s_\alpha\equiv\sin{\alpha}$.
As an example,
let us take $\rho-1 \in [0, 1] \times 10^{-3}$,
as in fig.~\ref{fig:1},
where the gray area corresponds to $\rho - 1 > 10^{-3}$.
We see that,
for small values of $\rho-1$, there is a large allowed region, including
very large values of $m_{Z_D}/m_Z$ for small values of $s_\alpha^2$.

\begin{figure}[t!]
  \centering
\includegraphics[width=0.75\textwidth]{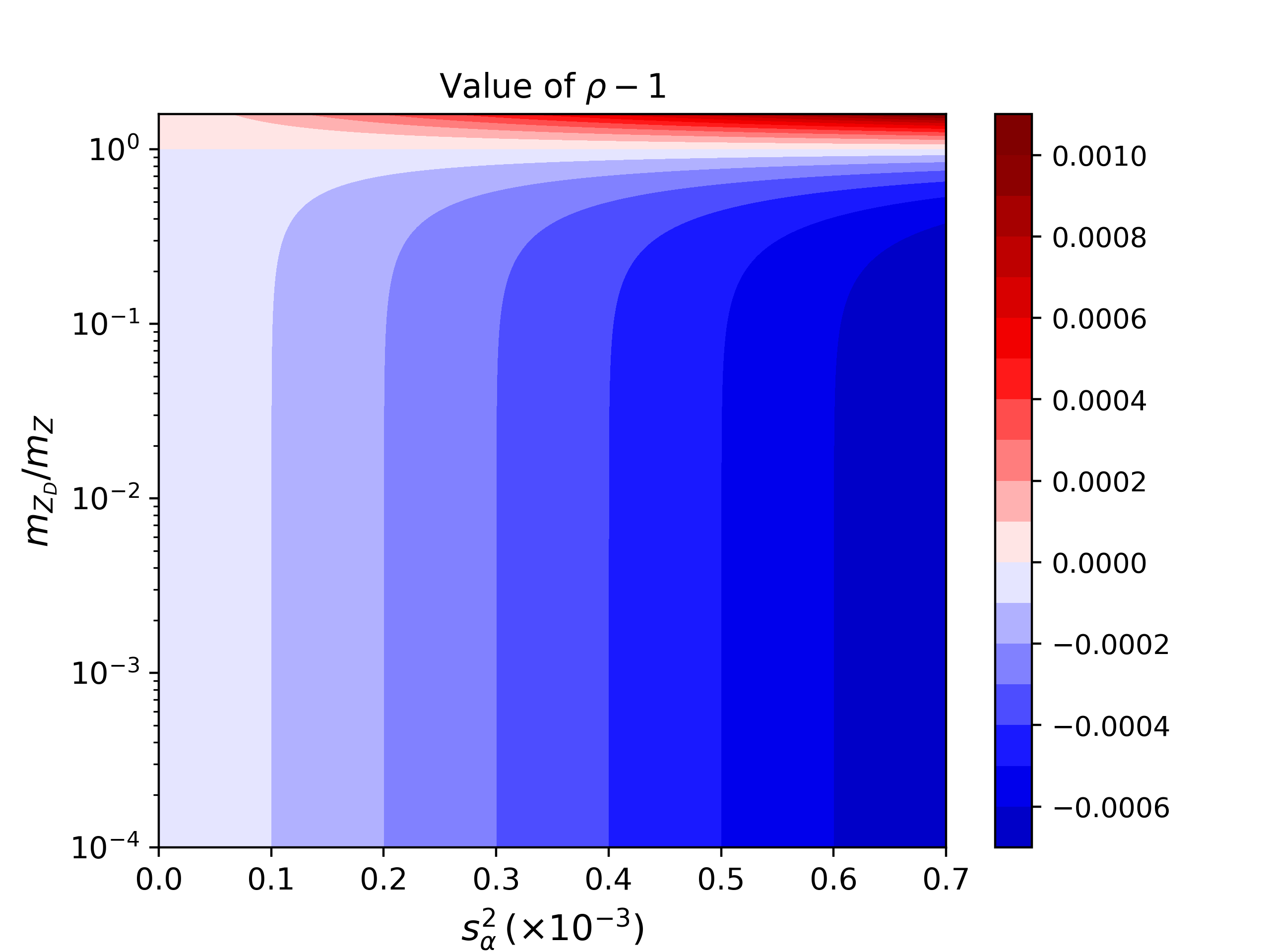}
\caption{Impact of \eq{master3} on the $(s_\alpha^2,\, m_{Z_D}/m_Z)$ plane,
for $\rho - 1 \in [-6, 10] \times 10^{-4}$.}
\label{fig:2}
\end{figure}

Fig.~\ref{fig:2} shows the same plane,
but allowing for negative values,
$\rho - 1 \in [-6, 10] \times 10^{-4}$.
We see that large values of $s_\alpha^2$ would only be possible 
if $m_{Z_D}$ were almost degenerate with $m_Z$.
In contrast,
very small values of $m_{Z_D}/m_Z$ require very small values for $s_\alpha^2$.

Although $\rho=1$ in the Standard Model, $\rho-1$ is negative
in the SU(2)$_L\times$U(1)$_Y\times$U(1)$_{Y'}$ model
if $m_{Z_D} < m_Z$.  For example,
when $\epsilon\ll 1$ and $1-\delta^2\sim\mathcal{O}(1)$, it follows that
\beq
\rho-1\simeq \frac{t_W^2\epsilon^2}{\delta^2-1}\,.
\label{aux334}
\eeq
Similarly,
in the SU(2)$_L\times$U(1)$_Y\times$U(1)$_{Y'}$ model,
$\rho-1$ is positive if $m_{Z_D} > m_Z$.

It is noteworthy that not all models with an extra $Z_D$ boson must have
$\rho \sim 1$ if $\epsilon \ll 1$, as explained in
appendix~\ref{app:p_legs}.
More generally, the elements of the squared mass matrix $\mathcal{M}_{ZZ_D}^2$ [see \eq{sqmassmat}] can be related to $\rho$ by using \eqst{mzzero}{mxz} and (\ref{master3}).
One then obtains:
\beqa
m_{Z^0}^2&=&\rho m_Z^2\,,\\[5pt]
m_X^2 &=&\left[ \frac{\cos^2 \alpha ( \rho - 1) + \sin^2 \alpha}{\rho - \cos^2 \alpha}\right] m_{Z_D^2}=\left(1+\frac{\rho-1}{\tan^2\alpha}\right)m_Z^2\,,\\
m_{XZ}^2&=&(1 - \rho) \frac{m_{Z}^2}{\tan \alpha} \, .\label{eq:rho_m12}
\eeqa
Any one of the above equations could serve as a definition of $\rho$ in 
an SU(2)$_L\times$U(1)$_Y\times$U(1)$_{Y'}$ model.   For example, employing
\eq{eq:rho_m12} in the model presented in section~\ref{gaugesector} yields:
\begin{equation}\label{eq:rho_m12_ours}
    \rho=\frac{1}{1-   \xi\eta t_W \epsilon \tan \alpha }\,.
\end{equation}
By measuring the $Z \bar{f} f$ interactions and separately determining a value of $\tan\alpha$,
one can extract a measurement of $\rho$ in the context of the dark-$Z$ model of section~\ref{gaugesector}.

All results presented in section~\ref{sec:unitarity_sum_rule}
involve tree-level parameters.  
In order to perform 
a more complete phenomenological study in which the parameters
of the model are constrained by precision electroweak observables, one must include 
the effects of radiative loop corrections.
For example, some of the dominant one-loop effects, which can be parameterized by the oblique parameters $S$, $T$, and $U$~\cite{Peskin:1991sw}, have
been incorporated in the analysis of the implications of generalized $Z$--$Z'$ mixing in ref.~\cite{Babu:1997st},
More recently, the one-loop radiative corrections to $m_W$
in an SU(2)$_L\times$U(1)$_Y \times$U(1)$_{Y'}$ model
have been examined in ref.~\cite{Peli:2023fyb}.  A more careful treatment of the radiative corrections to the results obtained in this section
and the resulting phenomenological consequences, which lie beyond the scope of the tree-level
analysis of this work, are currently under
investigation and will be reported in a future publication.

\subsection{Using unitarity as a consistency test}
\label{testing}

\enlargethispage{\baselineskip}

The model presented in section~\ref{gaugesector} has also been analyzed in
ref.~\cite{Foguel:2022unm}.  
In footnote~\ref{foot3}, we noted errors in two of the expressions given in ref.~\cite{Foguel:2022unm}.
One way to confirm the presence of such errors is by testing the model for consistency with unitarity, using
eq.~\eqref{eq:sum_rule}.

For example, eqs. (A.18)--(A.20) of ref.~\cite{Foguel:2022unm} exhibit the neutral vector boson squared masses and the mixing angle $\alpha$ under the assumptions that $|\epsilon|\ll 1$ and 
$\delta^2\ll 1$, 
\begin{align}
    &m_Z^2 \stackrel{\rm ?}{=}  m_{Z^0}^2 \left( 1 + \epsilon^2 t_W^2 \right) \, ,\label{eq:foguel_masses1} \\
    &m_{Z_D}^2 \stackrel{\rm ?}{=}  \delta^2 m_{Z^0}^2 \left( 1 - \epsilon^2 t_W^2 \right)  \,,\label{eq:foguel_masses2} \\
    & \sin^2\alpha  \stackrel{\rm ?}{=} \epsilon^2 t^2_W\,. \label{alphaapprox}
\end{align}
Consequently, \eqst{eq:def_parameters_sr1} {eq:def_parameters_sr5} yield:
\beqa
    &&g_{W^+ W^- \gamma}^2 = g^2 s_W^2 \, , \label{firsteq} \\
    &&g_{W^+ W^- Z}^2 \stackrel{\rm ?}{=}  g^2 c_W^2(1-\epsilon^2 t_W^2) \, ,\label{eq:foguelVVV1}\\
    &&g_{W^+ W^- Z_D}^2  \stackrel{\rm ?}{=}  g^2 c_W^2\,\epsilon^2 t_W^2\, ,\label{eq:foguelVVV2} \\
    &&g_{W^+ W^- h}^2 = g^2 m_W^2 c_h^2 \, ,  \\
    &&g_{W^+ W^- S^0}^2 = g^2 m_W^2 s_h^2 \,,\label{lasteq}
\eeqa
after employing \eq{alphaapprox}.
%
%
%
%
In the equations above, we have used the $\stackrel{\rm ?}{=}$ notation to indicate expressions 
that were 
obtained  using an inconsistent expansion in $\epsilon$ and~$\delta$ given
 in ref.~\cite{Foguel:2022unm}, where some terms of $\mathcal{O}(\epsilon^2\delta^2)$ 
were improperly discarded.\footnote{In footnote~\ref{foot3} we
noted that one should replace $\delta^2$ with $\eta^2\delta^2$ 
in the expression for $m^2_{Z_D}$ obtained in ref.~\cite{Foguel:2022unm}.   After expanding $\eta^2=(1-\epsilon^2/c_W^2)^{-1}$
around $\epsilon=0$, one derives an extra contribution of $\mathcal{O}(\epsilon^2\delta^2)$ to the expression for $m^2_{Z_D}$. 
This 
additional term, which is missing from \eq{eq:foguel_masses2},
 contributes terms of $\mathcal{O}(\epsilon^4\delta^2)$ to the violation of the unitarity sum rule given by eq.~\eqref{eq:sum_rule} and
to the violation of the relation $\rho^\prime=1$ 
 [after noting that $m_{Z_D}^2\sin^2\alpha\sim\mathcal{O}(\epsilon^2\delta^2)$].  These violations appear subsequently in the $\mathcal{O}(\epsilon^4)$ terms of
\eqs{eq:wrong_unitarity}{eq:wrong_rhoprime}, which we do not write out explicitly here.}
This inconsistency becomes evident when
evaluating the unitarity sum rule given by eq.~\eqref{eq:sum_rule} using the results of \eqst{firsteq}{lasteq},
which yields 
\begin{equation}\label{eq:wrong_unitarity}
    m_W^2-m_{Z^0}^2(c_W^2+s_W^2\epsilon^2\delta^2)+ \mathcal{O}\left(\epsilon^4\right)
    \stackrel{\rm ?}{=} 0 \, .
\end{equation}
%
As expected, the $\mathcal{O}(\epsilon^0)$ term in \eq{eq:wrong_unitarity} vanishes in light of \eq{mZ_def} [cf.~\eq{MZMWcW}].
However, a consistent expansion in $\epsilon$ and $\delta$ should yield exactly zero on the left-hand side of \eq{eq:wrong_unitarity} rather than 
nonzero terms of $\mathcal{O}(\epsilon^2)$ and $\mathcal{O}(\epsilon^4)$, respectively.

The impact of the inconsistent expansion in $\epsilon$ and $\delta$ can also be exhibited by computing~$\rho^\prime$, which was defined in \eq{rhoprime}.   Using \eqst{eq:foguel_masses1}{alphaapprox}, we obtain
\begin{equation} \label{eq:wrong_rhoprime}
    \rho' =\left(\frac{m_Z^2}{m^2_{Z^0}}\cos^2\alpha+\frac{m_{Z_D}^2}{m^2_{Z^0}}\sin^2\alpha\right)^{-1} \stackrel{\rm ?}{=}
    1 - \epsilon^2 t_W^2 \delta^2 + \mathcal{O}\left(\epsilon^4\right)
    \, ,
\end{equation}
which differs from the expected result of $\rho^\prime=1$ by a term of $\mathcal{O}(\epsilon^2)$.
Hence, we conclude that the expansions exhibited in \eqst{eq:foguel_masses1}{alphaapprox}, which were employed in
ref.~\cite{Foguel:2022unm}, violate tree-level unitarity.
In a correct analysis, \eqs{eq:foguelVVV1}{eq:foguelVVV2} should be replaced by \eqs{eq:VVV_expanded1}{eq:VVV_expanded2}, respectively.
Likewise, 
\eqs{eq:foguel_masses1}{eq:foguel_masses2}
should be replaced by \eqs{eq:masses_exp1}{eq:masses_exp2},
respectively, and \eq{alphaapprox} should be replaced by
\beq
\sin^2\alpha=\frac{\epsilon^2 t_W^2}{(1-\delta^2)^2}+\mathcal{O}(\epsilon^4)\,,
\eeq
in light of \eqs{Zmasseigen}{Xmasseigen}.
After making these substitutions, the unitarity sum rule given by \eq{eq:sum_rule} is restored, and the condition $\rho' = 1$ is satisfied exactly.

This exercise shows the power of the unitarity relations in probing the
consistency of a given model and/or the approximations chosen.

\section{\label{sec:concl}Conclusions}

The requirement of tree-level unitarity yields sum rules among the
couplings of a given theory.  The corresponding sum rules in electroweak models 
with an arbitrary scalar sector based on
the Standard Model gauge group SU(2)$_L \times $U(1)$_Y$ have been previously
obtained in refs.~\cite{Gunion:1990kf,Bento:2017eti,Bento:2018fmy}.
In this paper, we have
expanded the results given in the existing literature by considering models with an enlarged electroweak gauge group,
SU(2)$_L \times $U(1)$_Y \times $U(1)$_{Y'}$.
We have derived
sum rules involving gauge bosons and scalar bosons,
and we have obtained additional sum rules that include the couplings of gauge bosons and scalar bosons to fermions.
In particular,
we found an orthogonality of seemingly unrelated couplings,
extending results presented in ref.~\cite{Bento:2017eti,Bento:2018fmy} for multi-Higgs doublet models
with the Standard Model electroweak gauge group.

It is instructive to apply the unitarity sum rules obtained in this paper to a concrete extended electroweak model.  
Thus, we considered an
SU(2)$_L \times $U(1)$_Y \times $U(1)$_{Y'}$ gauge group which has been employed in the literature
to provide a model for a dark sector that consists of a new gauge boson (which has been called either a dark $Z'$ or
a dark photon)
that is feebly coupled to the Standard Model via kinetic mixing.   The dark $Z'$ can then be used to mediate
the interactions of a new fermion or scalar that is neutral with respect to the Standard Model gauge group and hence
is a candidate for dark matter.
In analyzing the SU(2)$_L \times $U(1)$_Y \times $U(1)$_{Y'}$ model described above, we have provided 
exact analytical results as well as approximate results that are obtained to first and second order in the kinetic mixing parameter. 
We then demonstrate how the unitarity sum rules can be used to provide consistency checks on these results (which allows
one to expose errors that have appeared in the literature due to inconsistent expansions).

Finally, we have introduced a parameter $\rho^\prime$ of the SU(2)$_L \times $U(1)$_Y \times $U(1)$_{Y'}$ model
that serves as the analog of the $\rho$ parameter of the Standard Model.  Whereas the tree-level value for $\rho$ in the
Standard Model is $\rho=1$, this latter result is modified in the SU(2)$_L \times $U(1)$_Y \times $U(1)$_{Y'}$ model,
whereas the tree-level value of $\rho^\prime=1$ is maintained.  In this analysis, it is important to define the tree-level
value of the weak mixing angle $\theta_W$ in terms of $m_W$, $\alpha_{\rm EM}$ and the Fermi constant $G_F$
in order to apply the same definition of $\theta_W$ to both the SU(2)$_L \times $U(1)$_Y$ and
SU(2)$_L \times $U(1)$_Y \times $U(1)$_{Y'}$ models.  Having done so, one can then relate
the definitions of $\rho$ and $\rho^\prime$,
to physical observables of the SU(2)$_L \times $U(1)$_Y \times $U(1)$_{Y'}$ model.   These results can be applied to 
models of dark photons,
such as those in refs.~\cite{Curtin:2014cca,Babu:1997st,Foguel:2022unm}, or to more specific studies of
dark matter, such as ref.~\cite{Mondino:2020lsc}.


\acknowledgments
\noindent

We are grateful to Bruno Bento, Ricardo Florentino, and Jorge C.~Rom\~{a}o for discussions.
The work of M.P.B. is supported in part by the Portuguese
Funda\c{c}\~{a}o para a Ci\^{e}ncia e Tecnologia\/ (FCT)
under contract SFRH/BD/146718/2019.
The work of M.P.B. and J.P.S. supported in part by FCT under Contracts
CERN/FIS-PAR/0008/2019,
PTDC/FIS-PAR/29436/2017,
UIDB/00777/2020,
and UIDP/00777/2020;
these projects are partially funded through POCTI (FEDER),
COMPETE,
QREN,
and the EU.
The work of H.E.H. is supported in part by the U.S. Department of Energy Grant No. DE-SC0010107.
H.E.H. is grateful for the hospitality and support during his visit to the Instituto Superior
T\'{e}cnico, Universidade de Lisboa.


\appendix

\section{\label{app:notations}Comparison of notation between this paper and others}

For the convenience of the reader, we provide a comparison of the
notation employed in section~\ref{sec:models} with that of
ref.~\cite{Babu:1997st} in table~\ref{table:parameters}.
%
\begin{table}[t!]
\centering
\begin{tabular}{| l || c | c |} 
 \hline
 Parameters & This paper & ref.~\cite{Babu:1997st}\TBstrut \\ [0.5ex] 
 \hline\hline
 sine of the weak angle & $s_W$  & $\hat{s}_W$\TBstrut \\ 
 \hline
 kinetic mixing term& $\epsilon$ & $- \hat{c}_W \sin \chi$ \TBstrut\\ 
 \hline
$X$ field rescaling factor& $\eta$ & $1/\cos \chi$ \TBstrut\\
 \hline
squared mass of $Z^0$ interaction-eigenstate & $m_{Z^0}^2$  & $\hat{M}_{Z}^2$\TBstrut \\
\hline
squared mass of $\hat{X}$ interaction-eigenstate & $\delta^2 m_{Z^0}^2$ &  $\hat{M}_{Z'}^2$\TBstrut \\
 \hline
 $Z^0$--$X$ squared mass mixing angle & $\alpha$ & $\xi$ \TBstrut \\
 \hline
\end{tabular}
\caption{Comparison between the parameters in this paper
and those of ref.~\cite{Babu:1997st}.}
\label{table:parameters}
\end{table}

To obtain the physical mass eigenstates of the neutral gauge boson, the first step is to perform a field redefinition
to obtain canonical kinetic energy terms (CK) for the neutral gauge fields.  One then constructs the $3\times 3$ squared mass matrix
of the neutral gauge bosons.  It is straightforward to identify the eigenstate with zero eigenvalue (corresponding to the photon), thereby
reducing the relevant neutral gauge boson squared mass matrix to a $2\times 2$ matrix.  In the final step, this matrix is diagonalized
to obtain the mass-eigenstate fields identified as the $Z$ boson of the SM and the dark $Z$ boson.  Schematically, the notation for the
fields used in this paper as compared to those of ref.~\cite{Babu:1997st} is indicated below:
\begin{align}
    & \text{This paper:} \quad \left( W^3, \hat{B}, \hat{X} \right)
    \xrightarrow{\rm CK} \left( W^3, B, X \right) \xrightarrow{\rm Photon}
    \left( A, Z^0, X \right) \xrightarrow{\rm Mass}
    \left( A, Z, Z_D \right) \, , \nonumber \\[5mm]
      & \text{ref.~\cite{Babu:1997st}:} \quad \left( W^3, \hat{B}, \hat{Z'} \right)
    \xrightarrow{\rm CK} \left( W^3, B, Z' \right) \xrightarrow{\rm Photon}
    \left( A, \hat{Z}, Z' \right) \xrightarrow{\rm Mass}
    \left( A, Z_1, Z_2 \right) \, .
\end{align}

Note that the model analyzed in ref.~\cite{Babu:1997st} is slightly
more general than the one considered here, since the former allows for the new scalar $S$ to have nonzero hypercharge~$Y$.
Setting the latter to zero corresponds to setting $\delta \hat{M}^2 = 0$ in the notation of 
ref.~\cite{Babu:1997st}.  Moreover,
ref.~\cite{Babu:1997st} defines one further field by 
\begin{equation}
    \hat{A} = \hat{s}_W W^3 +  \hat{c}_W \hat{B} \, .
\end{equation}
This definition is not needed in section~\ref{sec:models}, and thus we do not make use of it.

\section{\label{app:p_legs}Algebraic conditions for $\rho' = 1$}

The toy model introduced in section~\ref{sec:models} is one of many
examples of theories with a tree level value of $\rho' = 1$. 
More generally, consider a scalar extended SU(2)$_L\times$U(1)$_Y$  electroweak model, 
where each scalar multiplet (with weak isospin $T$ and U(1)$_Y$ hypercharge $Y$) satisfies
\begin{equation}\label{eq:t(t+1)}
    T(T+1) = 3 Y^2 \, .
\end{equation}
Such models naturally yield $\rho=1$ at tree-level, independently of the values of the neutral
scalar field vacuum expectation values.

If we extend the electroweak gauge group to SU(2)$_L\times$U(1)$_Y\times$ U(1)$_{Y^\prime}$, then we must
assign U(1)$_{Y'}$ quantum numbers to all the scalar fields.  As a concrete example, suppose we consider
a scalar sector that contains the SM Higgs doublet field and a second scalar multiplet with gauge quantum numbers $(T, Y, Y')$.
Then, the squared mass matrix previously obtained in \eq{sqmassmat} is modified as follows:
\begin{equation} \label{midifiedeq}
    \mathcal{M}^2_{Z Z_D} = m_{Z^0}^2
    \begin{pmatrix}
    1 - \frac{g^2 \delta^2 \left[ T(T+1) - 3 Y^2\right]}{2 c_W^2 g_D^2}  &\ \  \times \\
    \times &\ \  \times
    \end{pmatrix} \, ,
\end{equation}
where the matrix elements denoted by $\times$ above are not relevant for this discussion.  In particular, the matrix element explicitly
exhibited in \eq{midifiedeq} does \textit{not} depend on~$Y'$ and thus can be identified with  $m_{Z^0} = m_W / c_W$ when \eq{eq:t(t+1)} is satisfied.
Because the definition of $\rho'$ makes use of the relation
$m_{Z^0} c_W = m_W$ with $m_{Z^0}$ as defined in \eq{mZ_def}, the analysis of section~\ref{sec:a_new_rho} implies that $\rho' = 1$ at tree level
when \eqref{eq:t(t+1)} is satisfied, independently of the values of the neutral
scalar field vacuum expectation values.  

In the toy model introduced in section~\ref{sec:models}, the  SU(2)$_L\times$U(1)$_Y\times$ U(1)$_{Y^\prime}$ gauge quantum numbers of the scalar field $S$ are $(T,Y,Y')=(0,0,1)$. 
In this case, \eq{midifiedeq} reduces to the squared mass matrix
obtained in \eq{sqmassmat}, and we obtain $\rho^\prime=1$ as advertised.
Moreover, in this simple model,  the tree-level value of $\rho \sim 1$ [cf.~eq.~\eqref{aux334}]
because the off-diagonal element of \eq{sqmassmat}  were proportional to the small kinetic mixing parameter $\epsilon$.   This latter result does not persist in more general models where \eq{eq:t(t+1)} is satisfied and $\rho'=1$ is obtained.   To illustrate this point, consider a modification of the toy model 
examined in section~\ref{sec:models} in which $\epsilon=0$ and $S$ is replaced by a scalar field with gauge quantum numbers $(T,Y,Y')$ such that
\eq{eq:t(t+1)} is satisfied.  
In this case,  the squared mass matrix previously obtained in \eq{sqmassmat} is modified as follows: 
\begin{equation}
    \mathcal{M}^2_{Z Z_D} = m_{Z^0}^2
    \begin{pmatrix}
    1  &\ \ - \displaystyle\frac{g \, Y \, \delta^2}{c_W g_D} \\[12pt]
    - \displaystyle\frac{g \, Y\, \delta^2}{c_W g_D} &\ \  \delta^2
    \end{pmatrix} \, .
\end{equation}
Because this matrix is not diagonal,
$\sin \alpha$ is now determined by $v_D / v$ and $g_D$.  In this case, \eq{master3} shows that
the tree-level value of $\rho$ can deviate significantly from 1, since there is no reason for
$\sin\alpha$ to be particularly small.

\bibliographystyle{JHEP}
\bibliography{bibliography}

\end{document}